\documentclass{aa}  

\usepackage{graphicx}
\usepackage[varg]{txfonts}
\usepackage{upgreek}
\usepackage{subfig}
\usepackage{orcidlink}
\usepackage{amsmath}
\usepackage{textcomp}
\usepackage{orcidlink}
\usepackage{natbib,twoopt}
\bibpunct{(}{)}{;}{a}{}{,}           

\usepackage{hyperref}
\makeatletter
  \newcommandtwoopt{\citeads}[3][][]{\href{http://adsabs.harvard.edu/abs/#3}%
    {\def\hyper@linkstart##1##2{}
     \let\hyper@linkend\@empty\citealp[#1][#2]{#3}}}
  \newcommandtwoopt{\citepads}[3][][]{\href{http://adsabs.harvard.edu/abs/#3}%
    {\def\hyper@linkstart##1##2{}
     \let\hyper@linkend\@empty\citep[#1][#2]{#3}}}
  \newcommandtwoopt{\citetads}[3][][]{\href{http://adsabs.harvard.edu/abs/#3}%
    {\def\hyper@linkstart##1##2{}
     \let\hyper@linkend\@empty\citet[#1][#2]{#3}}}
  \newcommandtwoopt{\citeyearads}[3][][]%
    {\href{http://adsabs.harvard.edu/abs/#3}
    {\def\hyper@linkstart##1##2{}
     \let\hyper@linkend\@empty\citeyear[#1][#2]{#3}}}
\makeatother

\hypersetup{colorlinks=true,linkcolor=blue,citecolor=blue,urlcolor=blue,}
\makeatletter
\renewcommand*\aa@pageof{, page \thepage{} of \pageref*{LastPage}}
\makeatother
\begin{document} 

\title{Occurrence rate of stellar Type~II radio bursts from a 100\,star-year search for coronal mass ejections}

\author{David~C.~Konijn \orcidlink{0009-0000-7084-9956} \inst{1, 2}\fnmsep \href{mailto:konijn@astron.nl}{\thanks{Corresponding author: \texttt{konijn@astron.nl}}}
    \and
    Harish~K.~Vedantham \orcidlink{0000-0002-0872-181X} \inst{1, 2}
    \and
    Cyril~Tasse \inst{3, 4, 5}
    \and
    Timothy~W.~Shimwell \orcidlink{0000-0001-5648-9069} \inst{1, 6}
    \and \\
    Martin~J.~Hardcastle \orcidlink{0000-0003-4223-1117} \inst{7}
    \and
    Joseph~R.~Callingham \orcidlink{0000-0002-7167-1819} \inst{1, 8}
    \and
    Ekaterina~Ilin \orcidlink{0000-0002-6299-7542} \inst{1}
    \and
    Alexander~Drabent \orcidlink{0000-0003-2792-1793} \inst{9}
    \and \\
    Philippe~Zarka \orcidlink{0000-0003-1672-9878} \inst{4,10}
    \and 
    Floris~F.S.~van~der~Tak \orcidlink{0000-0002-8942-1594} \inst{11, 2}
    \and
    Sanne~Bloot~\orcidlink{0000-0002-3601-6165} \inst{1,2}
      }

\institute{
ASTRON, Netherlands Institute for Radio Astronomy, Oude Hoogeveensedijk 4, 7991 PD Dwingeloo, The Netherlands
    \and
Kapteyn Astronomical Institute, University of Groningen, Kapteynborg 5419, 9747 AD, Groningen, The Netherlands
    \and
LUX, Observatoire de Paris, Universit\'{e} PSL, CNRS, 5 Place Jules Janssen, 92190 Meudon, France
    \and
ORN, Observatoire de Paris, CNRS, PSL, Universit\'{e} d'Orl\'{e}ans, Nançay, France
    \and
Centre for Radio Astronomy Techniques and Technologies, Department of Physics and Electronics, Rhodes University, Grahamstown, 6140, South Africa
    \and
Leiden Observatory, Leiden University, PO Box 9513, 2300 RA, Leiden, The Netherlands
    \and
Centre for Astrophysics Research, Department of Physics, Astronomy and Mathematics, University of Hertfordshire, College Lane, Hatfield AL10 9AB, UK
    \and
Anton Pannekoek Institute for Astronomy, University of Amsterdam, Science Park 904, 1098 XH, Amsterdam, The Netherlands
    \and
Thüringer Landessternwarte, Sternwarte 5, D-07778 Tautenburg, Germany
    \and
LIRA, Observatoire de Paris, CNRS, PSL, Sorbonne Université, Université Paris Cité, Meudon, France
    \and
SRON, Netherlands Institute for Space Research, Landleven 12, 9747 AD, Groningen, The Netherlands
    }

\date{Accepted Astronomy \& Astrophysics}

% \abstract{}{}{}{}{} 
% 5 {} token are mandatory

\abstract{
Coronal mass ejections (CMEs) are major drivers of space weather in the Solar System, but their occurrence rate on other stars is unknown. A characteristic (deca-)metric radio burst with a time-frequency drift, known as a Type~II radio burst, is a key observational signature of CMEs. We searched a total of 107\,years of stellar data using time-frequency spectra that targeted all known stars within 100\,parsecs in the LOFAR Two Metre Sky Survey (LoTSS) up to May 2023. This resulted in the largest unbiased search for circularly polarised stellar Type~II metric radio bursts to date, with a typical $3\sigma$ sensitivity of $2.5$\,mJy for an integration time of 1\,minute. We detected two drifting stellar radio bursts: the published 2-minute burst from the M~dwarf StKM\,1-1262 and a new 13-minute burst from the M~dwarf LP\,215-56. The new burst is characterised by a drift rate of $-0.060^{+0.002}_{-0.002}$\,MHz\,s$^{-1}$, an average Stokes~V flux density of $-4.5^{+1.4}_{-1.3}$\,mJy, and a temporal duration of $63^{+31}_{-11}$\,seconds. We constrained the occurrence rate of drifting stellar bursts by calculating Poisson upper and lower limits based on the two drifting bursts. We also fitted a cumulative burst luminosity distribution to the data using the burst detections and the non-detections; this yielded a power law index ($\alpha$) of $-0.7^{+0.9}_{-0.6}$ and a normalisation point ($N$) of one burst per year with $E>6.8\times10^{13}$\,erg\,s$^{-1}$\,Hz$^{-1}$. We find an agreement between this and the cumulative luminosity distribution of decametric SOHO/LASCO solar Type~II data ($\alpha = -0.81 \pm 0.06 \pm 0.02$), which suggests that the current scarcity of detected stellar Type~II bursts is likely due to limited sensitivity rather than to the intrinsic rarity of these events. Additionally, we identify 19\,circularly polarised stellar radio bursts without a time-frequency drift.}

\titlerunning{Occurrence rate of stellar Type~II radio bursts from a 100\,star-year search for CMEs}

\maketitle

\section{Introduction}
Coronal mass ejections (CMEs) are violent eruptions of plasma from a stellar corona. On the Sun, they occur on average about once every few days\citepads{2024arXiv240704165G}. The bulk motion of the plasma, when super-Alfv\'enic, generates a shock that accelerates charges. The accelerated charges eventually cause radio emission via the plasma mechanism\citepads{1985ARA&A..23..169D}. The emission occurs at the local plasma frequency ($\nu_p$), given in Gaussian units as
\begin{equation}
    \nu_p = \left(\frac{n_e e^2}{\pi m_e}\right)^{1/2} \text{Hz,}
\label{eq:plasmaemission}
\end{equation}

\noindent where $n_e$ is the electron number density, and $m_e$ is the mass of an electron. 

As the super-Alfvénic shock propagates away from the star, the ambient plasma density decreases significantly from values typical of a corona (around $10^8 - 10^{10}$\,cm$^{-3}$). The emission frequency ($\nu_p \approx 100-1000$\,MHz at these densities) decreases, forming the characteristic time-frequency drift of a so-called Type~II radio burst (\citealt{paynescott1947,wild1950}). For the Sun, the connection between drifting radio bursts and CMEs has been confirmed by numerous coincident coronagraph and radio observations (e.g.\citeads{1972ARA&A..10..159W}; \citeads{1992ARA&A..30..113K}; \citeads{2012LRSP....9....3W}), with typical drift rates of $\sim-0.4$\,MHz/s at 150\,MHz (\citeads{2005ESASP.592..393A}) and burst durations of several minutes (\citealt{1959AuJPh..12..327R,1985srph.book..333N}).

Coronal mass ejections are major contributors to space weather and may influence planetary habitability. They can strip planetary atmospheres, affecting their long-term stability (e.g.\citeads{2007SSRv..129..207K};\citeads{2013SSRv..174..113L}), yet may also stimulate chemical reactions in planetary atmospheres that form prebiotic molecules (\citealt{2007AsBio...7...85S,2018ApJ...853...10L}). In particular, evidence suggests that the CME event rate or associated energies are elevated in certain classes of stars; if true, this could profoundly impact our understanding of the habitability of their exoplanets (e.g.\citeads{1997ApJ...480..344G}; \citeads{2009ApJ...703.2152T}).

To date, there has been no unequivocal detection of a stellar CME. Existing methods, such as observing extreme UV/X-ray coronal dimming (e.g.\citeads{2021NatAs...5..697V},\citeads{2022ApJ...936..170L},\citeads{2024ApJ...961...23N}), blueshifts in chromospheric lines (e.g.\citeads{2019A&A...623A..49V,2022NatAs...6..241N,2024ApJ...961..189N}), blueshifts in transition lines (e.g.\citeads{2011A&A...536A..62L}; \citeads{2019NatAs...3..742A}; \citeads{2024ApJ...969L..12I}), and X-ray/UV absorption (e.g.\citeads{1983ApJ...267..280H}, \citeads{2001ApJ...560..919B}), have revealed signatures consistent with stellar CMEs; however, we lack conclusive evidence that the plasma responsible for these signatures remains confined to the stellar magnetosphere or has dynamically decoupled from it. In contrast, Type~II radio bursts provide a more direct signature: they imply the existence of bulk plasma motion at super-Alfv\'{e}nic speeds. At these speeds the plasma cannot be confined by the stellar magnetic field, implying that the plasma producing the shock was ejected into interplanetary space.

Until recently, observational efforts to detect stellar Type II radio bursts had only yielded non-detections (e.g.\citeads{2018ApJ...862..113C};\citeads{2019ApJ...871..214V};\citeads{2020ApJ...905...23Z};\citeads{2021NatAs...5.1233C}). However, \citet{Callinghamsubmitted} and \citet{Tassesubmitted} have now reported the first detection of a stellar Type II burst, made possible by recent advances in low-frequency radio astronomy, particularly through the unprecedented sensitivity, frequency coverage, field of view, and resolution of the LOw-Frequency ARray (LOFAR;\citeads{2013A&A...556A...2V}), which enables the detection of more bursts and more stringent upper limits to be placed on their occurrence rate. Low-frequency observations (below $\sim$500\,MHz) are especially critical, as the anticipated plasma densities in stellar coronae will yield plasma emission at these frequencies (\citealt{2018ApJ...862..113C,2019ApJ...871..214V}).

In this work we present the largest and most comprehensive search for stellar Type~II radio bursts to date, identifying bursts with time-frequency drifts in time-frequency spectra (so-called dynamic spectra) synthesised by the post-processing code \texttt{DynSpecMS} \footnote{\texttt{https://github.com/cyriltasse/DynSpecMS}}\citep{Tassesubmitted}. \cite{Tassesubmitted} present the 233,899 time-frequency spectra for 93,608 unique stars within 100\,parsecs, derived from calibrated interferometric data in the LOFAR Two Meter Sky Survey (LoTSS;\citeads{2017A&A...598A.104S,2022A&A...659A...1S}). The 100-parsec boundary is motivated by detector sensitivity and computational constraints. At this distance, our derived 2.5\,mJy 3\,$\sigma$ sensitivity limit for 1-minute bursts corresponds to a radio luminosity of $3 \times 10^{16}$\,erg\,s$^{-1}$\,Hz$^{-1}$, several orders of magnitude brighter than observed solar Type~II bursts.

To improve detection sensitivity and reduce false positives, particularly those caused by radio frequency interference (RFI) and sidelobe structures of nearby bright sources, we searched for bursts via their circularly polarised emission (Stokes~V). Solar Type~II radio bursts are typically weakly circularly polarised (0-40 per cent; e.g.\citeads{1958AuJPh..11..201K};\citeads{1966AuJPh..19..209S};\citeads{1968SvA....12...21F}), although instances of strongly circularly polarised emission are theoretically plausible\citepads{2003ApJ...592.1234T}. While Stokes~V data reduce our sensitivity to weak circularly polarised emission, they significantly enhance detection reliability. This is because Stokes~V is far less affected by the sidelobe structures of other sources compared to the total intensity, given that Stokes~V sources are inherently scarcer (\citealt{2018MNRAS.478.2835L,2023AA...670A.124C}).

In this paper we address whether stellar Type~II radio bursts occur as frequently as their solar counterparts. We begin by detailing our observational setup and data reduction strategy in Sect.~\ref{sec:obs}. We then describe our methodology for identifying and characterising potential burst candidates in Sect.~\ref{sec:searchbursts}. In Sect.~\ref{sec:burst properties} we analyse the properties of the detected stellar bursts and establish upper limits on the Type~II burst rate. Finally, we interpret our findings in Sect.~\ref{sec:disc} and summarise our key conclusions in Sect.~\ref{sec:conc}.
\section{LoTSS data and post-processing} \label{sec:obs}
The methodology used to generate the dynamic spectra used in this paper is described in \citet{Tassesubmitted}; here we summarise the survey characteristics for context: LoTSS is an ongoing survey of the northern sky through eight-hour exposures each of 3168 pointings at 120 to 168\,MHz\citepads{2017A&A...598A.104S}. The dataset encompasses all data processed up to May 2023, comprising 1651 pointings, which totals a combined coverage of 12,500\,deg$^2$. This includes observations from both LoTSS Data Release 1 and Data Release 2, as well as additional data beyond these releases, forming part of the upcoming Data Release 3 (DR3).
The standard LoTSS pipeline\footnote{\texttt{https://github.com/mhardcastle/ddf-pipeline}} produces high-fidelity, thermal noise-limited images and source-catalogues using direction-dependent self-calibration (\citealt{2014arXiv1410.8706T,2015MNRAS.449.2668S,2018A&A...611A..87T,2021A&A...648A...1T}). With \texttt{DynSpecMS}, the clean source components are subtracted from the visibilities using their direction-dependent gains. The residual visibilities are then coherently beam-formed towards our stellar targets to produce full-Stokes dynamic spectra at the native time--frequency resolution of the processed visibilities (8\,seconds and $\sim$100\,kHz). Per pointing, $\sim$100 dynamic spectra are also generated on `blank-sky' locations near the stellar targets, away from any radio sources, to cross-check residual systematic errors from faint interference and calibration issues. A typical LoTSS pointing contains $\sim$\,100\,stars within 100\,pc, resulting in $100\times 8\,{\rm hr}$ worth of stellar dynamic spectra data. In total, 93,608\,unique stellar objects were observed, each for approximately 8\,hours, though some pointings were observed multiple times.

The full observational campaign totals 1,264,977.74\,hours, or 144.31\,years. However, this total exposure time includes stellar objects like white dwarfs, which are unlikely to host CMEs. To calculate a more representative exposure time, we refined our \textit{Gaia} sample by applying specific data quality filters: we included all \textit{Gaia} DR3 stellar objects within 100\,pc\citepads{2023A&A...674A...1G} and we adopted a 10 per cent relative precision criterion: \texttt{parallax\_over\_error > 10}. Similarly, we applied filters to the relative flux error on the $G_{BP}$, and $G_{RP}$ photometry: \texttt{phot\_rp\_mean\_flux\_over\_error > 10}, and \texttt{phot\_bp\_mean\_flux\_over\_error > 10}. We included only stars with high-quality astrometric solutions: \texttt{visibility\_periods\_used > 6}. Finally, as outlined by \cite{2018A&A...616A..10G}, we used an additional filter to limit our analysis to sources within the empirically defined locus of the $I_{BP} + I_{RP}/I_G$ fluxes ratio as a function of $G_{BP}-G_{RP}$ colour: \texttt{phot\_bp\_rp\_excess\_factor > 1.0+0.015} $(G_{BP}-G_{RP})^2$ and \texttt{phot\_bp\_rp\_excess\_factor < 1.3+0.06} $(G_{BP}-G_{RP})^2$.  

\begin{figure}
    \centering
    \includegraphics[width=1\linewidth]{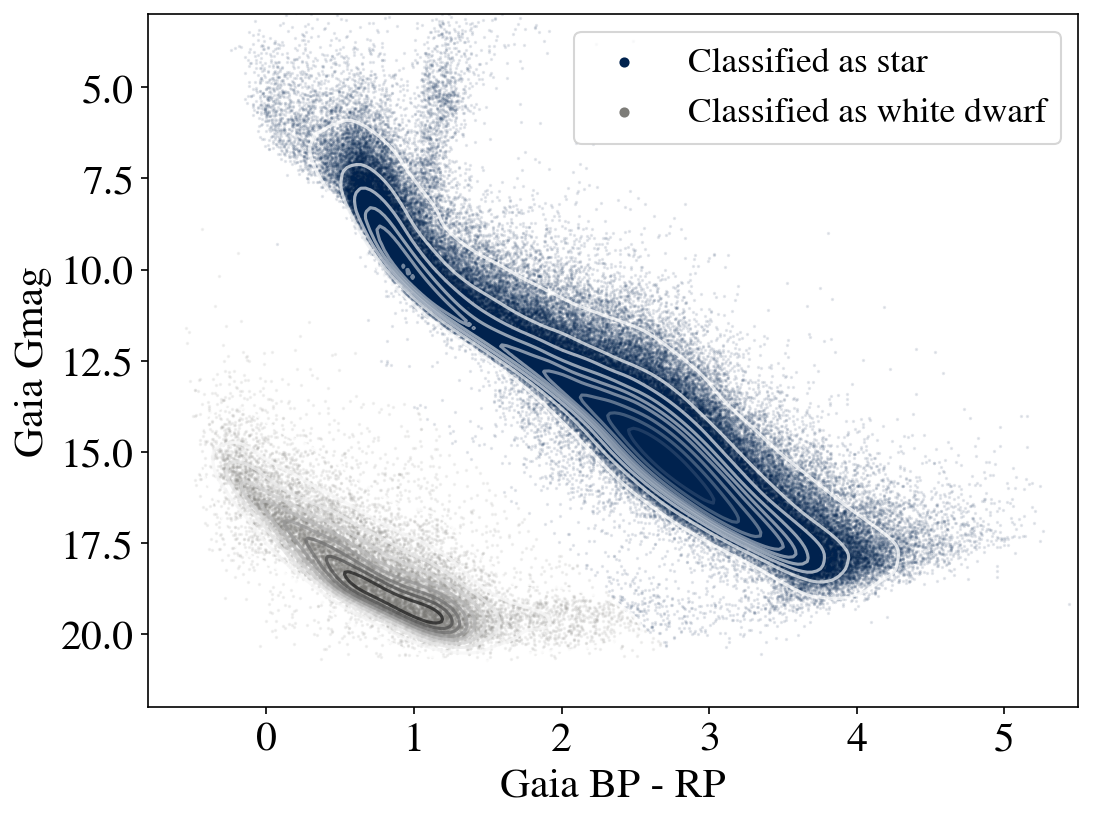}
    \caption{Colour-magnitude diagram of \textit{Gaia} sources observed with LoTSS, applying the quality filters described in Sect.~\ref{sec:obs}. The sources are categorised into two clusters based on their proximity in the colour-magnitude space.}
    \label{fig:HR_sample_lotss}
\end{figure}

We removed white dwarfs from the total sample by iteratively grouping data points into two distinct clusters based on their proximity in the colour-magnitude space and then excluding the cluster corresponding to white dwarfs. The final sample, shown in Fig.~\ref{fig:HR_sample_lotss}, focuses specifically on potential CME hosts and has been observed for a total of 107.21\,years. Note that we processed and analysed the entire 144.31\,years of data, but we used the 100-year duration when calculating total CME event rates.

\section{Search for and discovery of bursts} \label{sec:searchbursts}
We searched for drifting bursts by analysing the Stokes~V time series data. We averaged over the frequency axis after correcting for different trial sweep rates. To mitigate the effects of residual RFI, the frequency averaging uses inverse-variance weighting, where each data point is weighted by the inverse of its blank-sky variance.

\subsection{Inverse-variance weighted time series}
The inverse-variance weighted average ($\hat{y_t}$) at a given time ($t$), after drift-correction, is calculated as\begin{equation} 
\hat{y_t} = \frac{\sum_i y_i/\sigma^2_i}{\sum_i 1/\sigma^2_i}, 
\end{equation} 
\noindent where $y_i$ is the flux density of the target frequency channel ($i$), and $\sigma^2_i$ is the variance for frequency channel $i$ at time $t$, which is calculated using the blank-sky spectra as\begin{equation}
\sigma^2_i = \frac{\sum_j w_j (b_{j,i} - \bar{b}_i)^2}{\sum_j w_j},
\end{equation}
\noindent where $b_{j,i}$ is the flux density of blank-sky spectrum $j$ in frequency channel $i$, $w_j$ is the weighting factor for blank-sky spectrum $j$, and $\bar{b}_i$ is the weighted mean:
\begin{equation}
\bar{b}_i = \frac{\sum_j w_j b_{j,i}}{\sum_j w_j}.
\end{equation}
\noindent Due to RFI and calibration errors, some blank-sky pixels appeared excessively bright, causing inverse-variance weighting to down-weight the target pixel, even when it was unaffected. To address this, we implemented several strategies to adjust the contribution of each individual blank-sky spectrum so that the combined $\sigma^2_i$ best represents the target data:
\begin{equation}
w_j = \begin{cases}
\frac{N_\text{off-facet} - N_\text{in-facet}}{N_\text{in-facet}} & \text{if spectrum } j \text{ is in-facet} \\
0 & \text{if spectrum } j \text{ is in top 10\% outliers} \\
1 & \text{otherwise (off-facet, non-outlier).}
\end{cases}
\end{equation}
With this ancillary weighting and flagging, we obtained an inverse-variance weighted time series in which RFI is largely suppressed.

\subsection{Drift-corrected search}
We constructed the frequency averaged time series by applying a range of trial drift corrections to the dynamic spectrum, effectively compensating for frequency-dependent time delays. To achieve this, we adopted a simplified drift model in which a shock propagates at a constant speed through a medium where the density ($n$)  decreases as $n \propto \frac{1}{r^2}$, where $r$ is the radial distance from the star. Since the plasma frequency follows $\nu_p \propto n^{1/2}$, this implies $\nu_p \propto 1/r$. We can express the radial position of a shock moving at constant radial speed as $r = r_0 + vt$, where $r_0$ is the starting position, $v$ the speed of the shock, and $t$ the time. This leads to

\begin{equation} 
\frac{d\nu_p}{\nu_p} = -\frac{dr}{r} = -\frac{v}{r} dt.
\end{equation}

\noindent This in turn gives

\begin{equation} 
\frac{dt}{d\nu_p} = -\frac{C}{v \nu_p^2}, 
\end{equation}

\noindent where $C$ is a constant, arising from $\nu_p = C/r$. If we integrate from an initial frequency $\nu$ to $+\infty$, we obtain the arrival time:

\begin{equation} 
t(\nu_p) = \frac{C}{v \nu_p}.
\end{equation}

\noindent The time delay across our observing band is therefore

\begin{equation}
\Delta t = \frac{C}{v \nu_{\text{min}}} - \frac{C}{v \nu_{\text{max}}}
\label{eq:delay}
.\end{equation}
Using this time-delay model, similar to pulsar de-dispersion techniques, we could `unsweep' the drift of the burst in the dynamic spectrum, counteracting the delay on a per-channel basis and aligning the burst vertically in the dynamic spectrum. Next, we computed the inverse-variance weighted time series on these drift-corrected spectra, enhancing the S/N for a drift that effectively aligns the burst vertically. The drift value that achieves the best vertical alignment, i.e. the highest S/N, represents the true drift of the burst. 

\begin{figure*}
    \centering
        \includegraphics[width=0.49\textwidth]{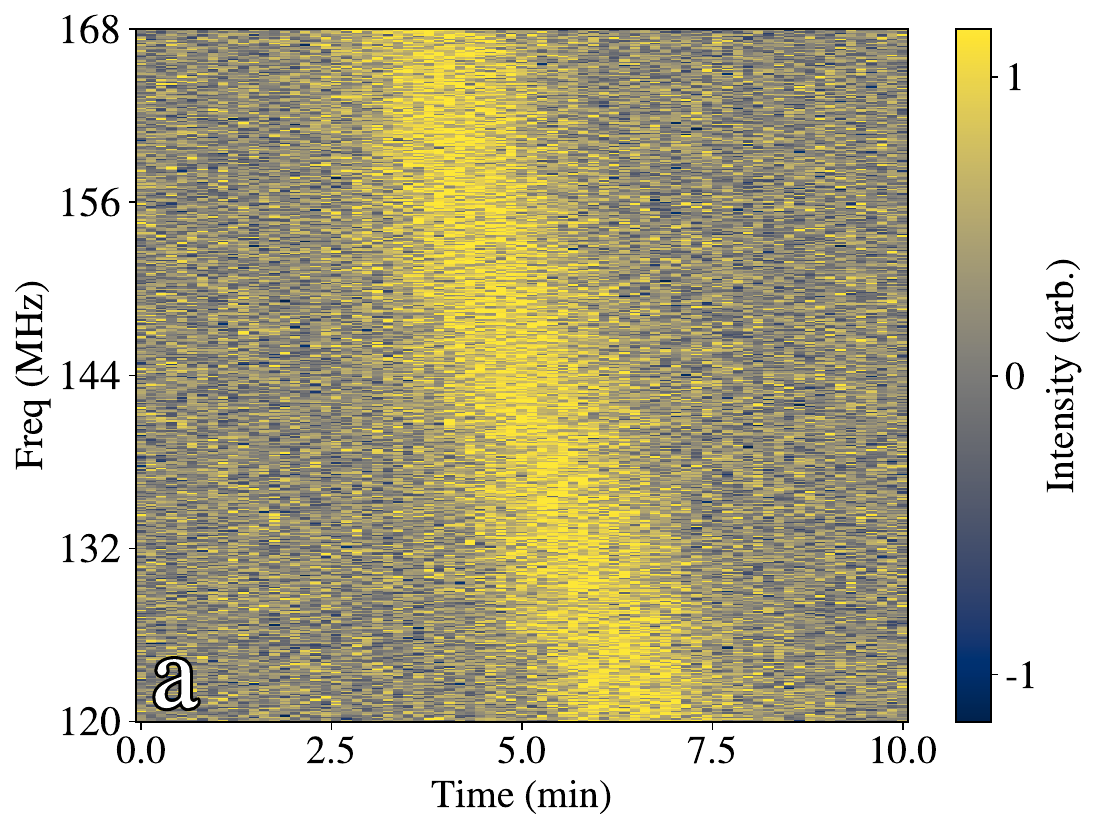}
    \hfill
        \includegraphics[width=0.49\textwidth]{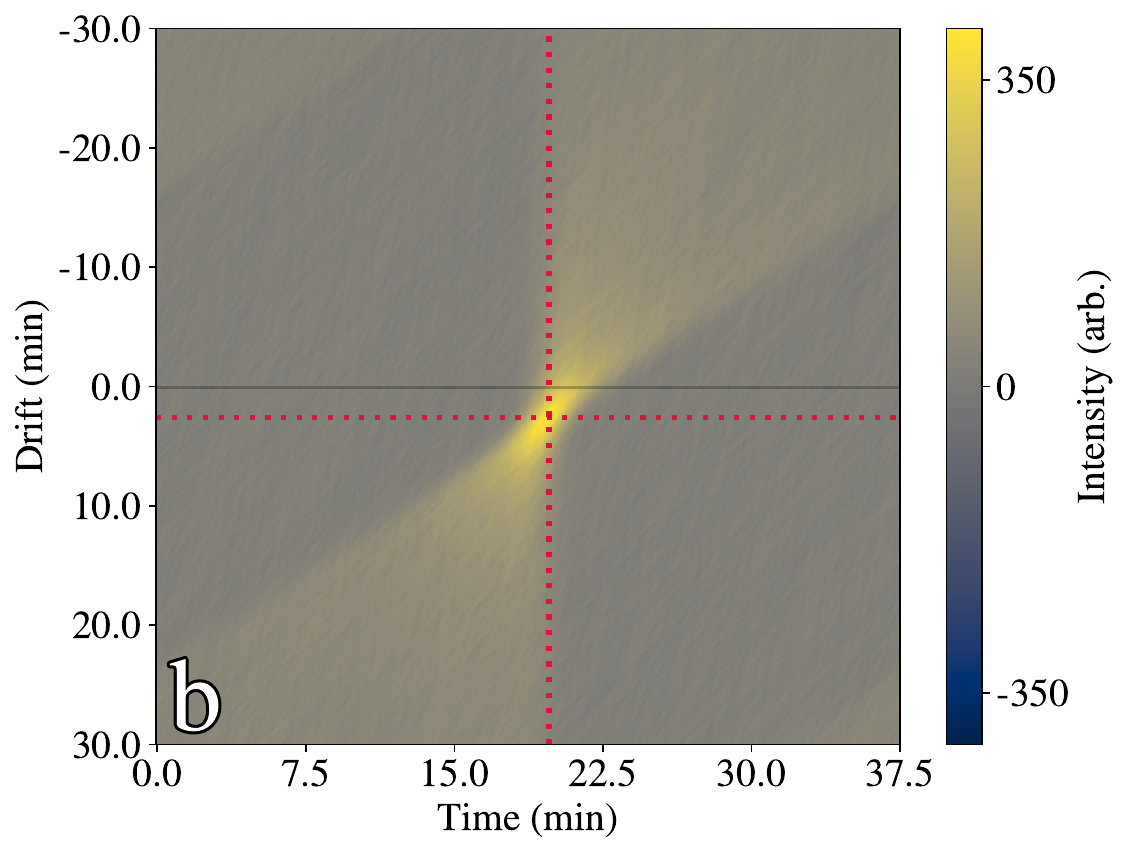}
    \caption{Panel~a: Mock Stokes~V dynamic spectrum of a radio burst due to plasma emission modelled using the Parker Solar wind model with solar parameters, assuming a shock speed of 1000\,km\,s$^{-1}$ and a S/N of 30. Panel~b: Frequency-integrated burst brightness as a function of the drift delay and time, modelled using the Parker Solar wind model with solar parameters, assuming a shock speed of 1000\,km\,s$^{-1}$. The defining `bow-tie'-shaped feature of a broadband swept astrophysical signal is clearly visible. The red cross indicates the brightest pixel and its corresponding burst drift.}
    \label{fig:parker}
\end{figure*}

We searched for bursts by setting the range of trial drifts to extend up to 30\,minutes, as Type~II bursts do not often last longer \citepads{2023A&A...675A.102K}. Using a range of trial drifts up to 30 minutes results in 225 times the original number of time series (30\,minutes at an 8-second resolution), each corrected for progressively larger drifts. We applied a boxcar matched-filter search with widths ranging from 8 to 600\,seconds to each drift-corrected time series, searching for events with S/N$\geq$8 to minimise false-positive detections that are generated by Gaussian noise. Our data consist of $2.73\times10^{11}$ data points, which we sampled an additional $225$\,times with our drift-corrected search. Gaussian noise is expected to produce an 8-sigma event approximately once every $1.61\times10^{15}$ samples, resulting in fewer than one false-positive detection due to Gaussian noise. Since the noise in our data may not be perfectly Gaussian, an 8-sigma threshold should at least significantly reduce false detections.

To test if our new method can reliably detect drifting bursts in the dynamic spectrum, we simulated a drifting burst by modelling a stellar density profile using the Parker Solar wind model \citepads{1965SSRv....4..666P}, through which a shock propagates (see Appendix~\ref{appendix:parker}). As the shockwave passes through the wind and emits via plasma emission, the emission frequency will be directly related to the density profile. Assuming a base wind electron density of $10^{9}$\,cm$^{-3}$, a coronal temperature of 3\,$MK$, and a fractional bandwidth of 0.25 \citepads{2005JGRA..11012S08A}, a shockwave moving at a constant velocity of 1000\,km\,s$^{-1}$ could produce a signal in our dynamic spectra as shown in Fig.~\ref{fig:parker}a. We defined the sign of the Stokes~V emission as left-hand circularly polarised light minus right-hand circularly polarised light (\citeads{2021A&A...648A..13C}; \citeads{2021NatAs...5.1233C}). The drift-corrected time series can be visualised in a `drift-time' spectrum, which shows the inverse-variance weighted frequency-integrated burst brightness as a function of drift and time. For the Parker Solar wind burst, this is illustrated in Fig.~\ref{fig:parker}b. In this parameter space, any broadband signal, whether drifting or not, gives rise to a symmetric `bow-tie'-like shape. 

To further confirm that the detected bursts are of stellar origin and are not created by RFI, we created Stokes~V images with a resolution of 20$"$ using \texttt{WSClean} \citep{2014ascl.soft08023O, 2014MNRAS.444..606O}, where we imaged along the identified drift; only using the data within the time periods covered by the drift, flagging the rest. From the image, we can determine if the source of the emission originates from a point source in the sky at the location of the star. However, this imaging process is computationally expensive, especially for a large number of burst candidates. Therefore, we first manually inspected each burst to reject candidates created by RFI or sidelobe patterns of nearby bright sources, which often exhibit features such as negative total intensity burst emission or non-physical, rapidly changing intensities across both time and frequency channels. We also rejected any candidate for which the Stokes~I emission is less than half of the emission in Stokes~V. Only after passing these checks did we image the burst to identify if the emission originates from a point source.

\section{Burst properties} \label{sec:burst properties}
\begin{figure*}
    \centering
    \includegraphics[width=0.93\linewidth]{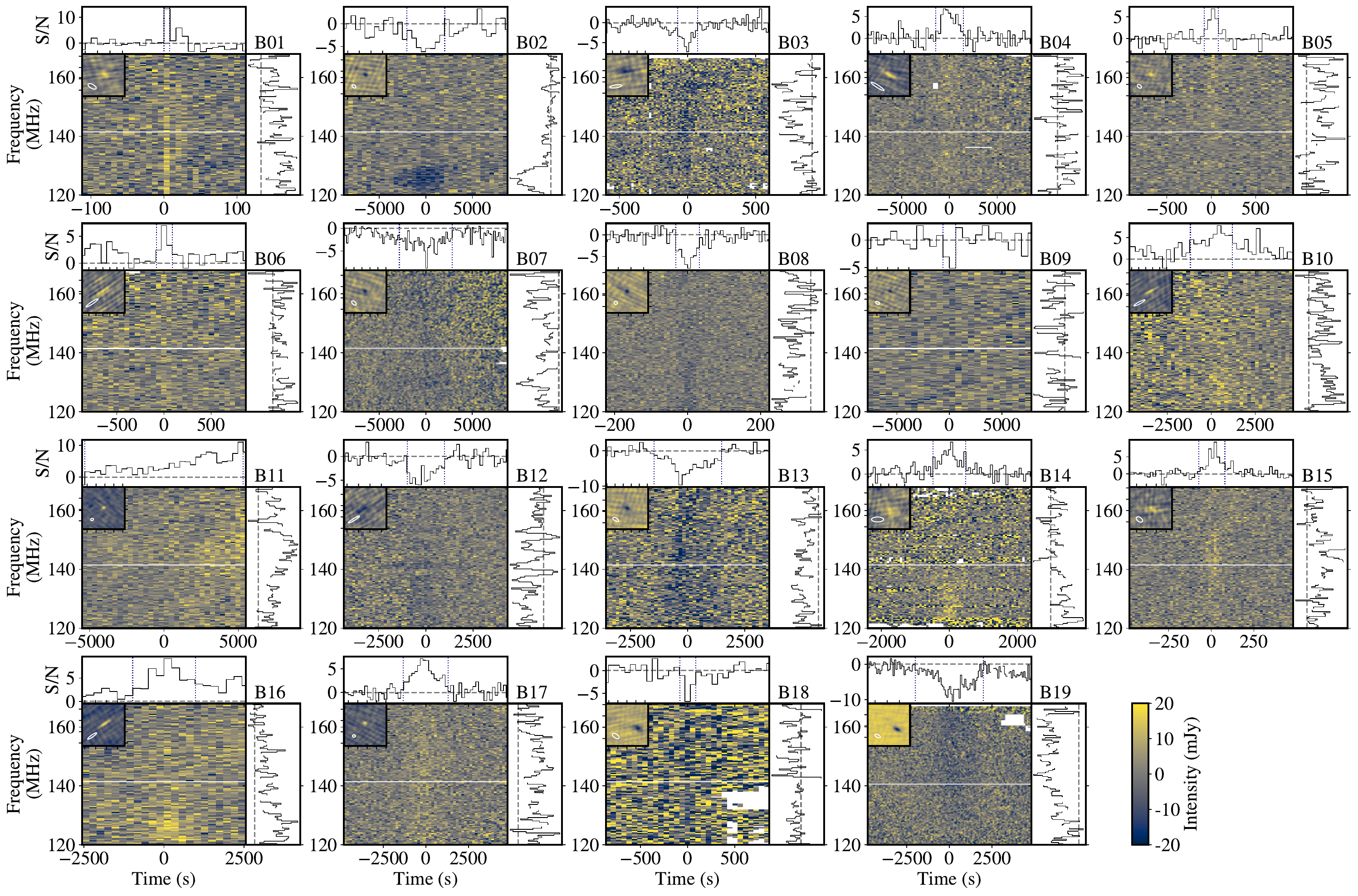}
    \caption{The 19 stellar radio bursts detected without drifting features. Each thumbnail shows a dynamic spectrum ($\Delta\nu$\,=\,0.4\,MHz, $\Delta$t\,=\,the boxcar width). The interferometric images are shown as insets in the top left. The time series of the inverse variance weighted frequency-integrated spectra are shown at the top of each panel, with vertical dotted lines indicating the start and end times of the burst. The time-integrated spectra (shown to the right of each panel) are integrated over the time range indicated by the  vertical dotted blue lines in the time series. Denoted in the top right of each thumbnail is the burst ID, which corresponds to Table~\ref{tab:hosttable}. White areas in the dynamic spectra represent areas masked due to the presence of RFI, and the white ovals in the interferometric image represent the point spread function. For visual purposes, the colour maps of the dynamic spectra range from -20 to 20\,mJy. Bursts B02, B04, B10, B11, B12, B14, B16, B17, and B20 are also presented in \citet{Tassesubmitted}.}
    \label{fig:family bursts}
\end{figure*}

With our drift-search method, we can also identify bursts with zero drift, as their S/N will peak at zero drift-correction. Using our approach, we detected 21 stellar radio bursts: 19 without an observable drift, and 2 with drifting features. Table~\ref{tab:hosttable} shows the stellar hosts and the time of arrival for the 21\,detected stellar bursts. Additionally, to provide an indication of the stellar activity levels, we included literature values of X-ray fluxes converted to X-ray luminosities for each star, as CMEs are often associated with enhanced X-ray emission\citepads{1975SoPh...45..377S}. Where published X-ray fluxes were unavailable, we estimated upper limits using data from the ROentgen SATellite (ROSAT; \citeads{1982AdSpR...2d.241T}) All-Sky Survey\citepads{1999A&A...349..389V}. Figure~\ref{fig:family bursts} shows the dynamic spectra of the stellar bursts without any drifting features, where the spectra were smoothed to optimise clarity. The two drifting bursts we detect are shown in Figs.~\ref{fig:joesburst} and~\ref{fig:13minburst}. Burst B20 corresponds to the event reported by \citet{Callinghamsubmitted}.

\begin{table*}
\caption{Stellar hosts and times of arrival for the 21\,detected stellar bursts.}
\label{tab:hosttable}
\renewcommand{\arraystretch}{1.25}
\begin{tabular*}{\textwidth}{@{\extracolsep{\fill}} l l l l l l l}
\hline \hline
Burst ID            & Obs. ID & Source Name$^{\text{a}}$    & Spectral Type         & X-ray Luminosity$^{\text{b}}$     & Distance          & Time of Arrival$^{\text{c}}$ \\
                    &         &                             &                       & (erg\,s$^{-1}$)                   & (pc)              & (MJD) \\\hline
B01                 & L695587 & LP 192-44                   & M6V$^{\text{g}}$      & $<$\,$9.52\times 10^{28}$         & 53.30             & 58534.67984606 \\
B02$^{\text{d}}$    & L800676 & G 250-25                    & M3.5V$^{\text{h}}$    & $(1.58 \pm 0.52) \times 10^{29}$  & 30.03             & 59188.19808440 \\
B03                 & L798434 & PM J07349+1445              & M3V$^{\text{h}}$      & $(4.64 \pm 0.60) \times 10^{29}$  & 16.30             & 59175.18451620 \\
B04$^{\text{d}}$    & L664734 & PM J09190+4610              & M3.5V$^{\text{i}}$    & $(1.32 \pm 0.27) \times 10^{29}$  & 30.03             & 58351.72629861 \\
B05                 & L727610 & G 236-2                     & M3.5V$^{\text{h}}$    & $(1.09 \pm 0.13) \times 10^{28}$  & 11.86             & 58668.82176042 \\
B06                 & L665242 & LP 261-57                   & M4V$^{\text{i}}$      & $<$\,$3.11\times 10^{29}$         & 81.82             & 58362.39292940 \\
B07                 & L728721 & AD Leo                      & M3V$^{\text{h}}$      & $(1.95 \pm 0.09) \times 10^{29}$  & 4.97              & 58677.81698611 \\
B08                 & L258201 & G 123-35                    & M3.5V$^{\text{h}}$    & $(1.65 \pm 0.21) \times 10^{29}$  & 23.51             & 57044.53220486 \\
B09                 & L612032 & LSPM J1228+6515             & M3.5V$^{\text{i}}$    & $<$\,$1.34\times 10^{29}$         & 71.57             & 58023.74614583 \\
B10$^{\text{d, e}}$ & L400133 & 2MASS J14333139+3417472     & M5V$^{\text{i}}$      & $<$\,$1.12\times 10^{29}$         & 48.21             & 57305.49466319 \\
B11$^{\text{d, e}}$ & L603984 & G 223-74                    & M3V$^{\text{h}}$      & $(6.61 \pm 0.73) \times 10^{28}$  & 18.46             & 57982.14790972 \\
B12$^{\text{d, e}}$ & L259433 & i Boo                       & G9/F5V$^{\text{j}}$   & $(1.84 \pm 0.03) \times 10^{30}$  & 12.95             & 57058.28746528 \\
B13                 & L763737 & LP 443-17                   & M5V$^{\text{h}}$      & $<$\,$1.38\times 10^{28}$         & 17.10             & 58845.44449653 \\
B14$^{\text{d}}$    & L746965 & BD+19 5116B                 & M4V$^{\text{h}}$      & $(2.90 \pm 0.09) \times 10^{29}$  & 6.25              & 58753.94009722 \\
B15                 & L719904 & HD 147379B                  & M3V$^{\text{h}}$      & $(2.76 \pm 0.29) \times 10^{28}$  & 10.76             & 58636.10437963 \\
B16$^{\text{d}}$    & L652062 & G 240-45A                   & M4V$^{\text{h}}$      & $<$\,$8.61\times 10^{27}$         & 27.49             & 58240.94514236 \\
B17$^{\text{d}}$    & L857710 & BD+68 946                   & M3V$^{\text{h}}$      & $(5.35 \pm 0.30) \times 10^{27}$  & 4.55              & 59687.44707176 \\
B18                 & L694975 & HD 239960A                  & M3V$^{\text{h}}$      & $(6.02 \pm 0.54) \times 10^{27}$  & 4.01  & 58521.61115394 \\
B19                 & L696951 & EV Lac                      & M4V$^{\text{h}}$      & $(1.13 \pm 0.03) \times 10^{29}$  & 5.05  & 58550.55276620 \\
B20$^{\text{d, f}}$ & L470106 & StKM 1-1262                 & M0V$^{\text{f}}$      & $(3.80 \pm 0.32) \times 10^{29}$  & 40.88             & 57525.92432407 \\
B21                 & L695755 & LP 215-56                   & M2.4V$^{\text{h}}$    & $<$\,$5.59\times 10^{29}$         & 62.13             & 58552.21916088 \\
\hline \hline
\multicolumn{7}{l}{$^{\text{a}}$ See \texttt{https://simbad.cds.unistra.fr/guide/otypes.htx}}\\
\multicolumn{7}{l}{$^{\text{b}}$ X-ray catalogue flux values are taken from \citetads{2017MNRAS.469.1065D} when available (0.1--2.4 keV for ROSAT, 0.2--12 keV for XMMSL);} \\
\multicolumn{7}{l}{\phantom{$^{\text{b}}$} otherwise, detections or upper limits were derived using the \texttt{XMM Upper Limit Server}$^{\textcolor{blue}{3}}$ for the ROSAT mission (0.2--2 keV),} \\
\multicolumn{7}{l}{\phantom{$^{\text{b}}$} 
 assuming a power law model $\Gamma=2$, and  N$_{\text{H}}=10^{20}$ cm$^{-2}$.} \\
\multicolumn{7}{l}{$^{\text{c}}$ Uncertainties of $\sim$\,4\,seconds.} \\
\multicolumn{7}{l}{$^{\text{d}}$ These bursts are presented in detail by \citet{Tassesubmitted}.} \\
\multicolumn{7}{l}{$^{\text{e}}$ These bursts are presented in detail by \citetads{2023AA...670A.124C}.} \\
\multicolumn{7}{l}{$^{\text{f}}$ This burst is presented in detail by \citet{Callinghamsubmitted}.} \\
\multicolumn{7}{l}{$^{\text{g}}$ Spectral types based on effective temperature }\\
\multicolumn{7}{l}{$^{\text{h}}$ Spectral types adopted from \citetads{2000AAS..143....9W}.}\\
\multicolumn{7}{l}{$^{\text{i}}$ Spectral types adopted from \citetads{2016MNRAS.457.2192C}.}\\
\multicolumn{7}{l}{$^{\text{j}}$ Binary system, unknown which star is the host.} \\

\end{tabular*}
\end{table*}

\subsection{Drift characterisation}
To obtain statistical estimates for the drift parameters, we applied a Markov chain Monte Carlo (MCMC) algorithm to fit a drift model to the data, which can be described by four parameters: the constant burst flux density $S_0$, the arrival time $t_0$, the drift rate $a$, and the burst width $\Delta t$. We again modelled the time delay as Eq.~\ref{eq:delay}. A drifting burst model takes the form

\begin{equation}
    S_k = S_0 W\left( t_0 + \frac{a}{\nu_k}, \Delta t \right), 
\end{equation}

\noindent where $k$ is the frequency channel number and $W$ is given by

\begin{equation}
    W(t, \Delta t) =
        \begin{cases}
        1 & ; -\Delta t/2 < t < \Delta t/2\\
        0 & ; \text{otherwise.}
    \end{cases}   
\end{equation}

\addtocounter{footnote}{3}
\footnotetext[3]{\url{http://xmmuls.esac.esa.int/upperlimitserver}} 

\noindent The MCMC approach maximises the sum of the log-likelihood function ($\mathcal{L}$), defined by

\begin{equation}
   \mathcal{L} = C - \sum_k \frac{| \text{data}  - \text{model}|^2}{2 \sigma^2_k},
\end{equation}

\noindent where $C$ is a constant, $k$ a specific frequency channel, and $\sigma^2_k$ the variance in that channel. For our drift model, the likelihood is expressed as

\begin{equation}
    \text{Likelihood}  = \prod_k \exp{({-[S_k - S_0 W(t_k + t_0 + \frac{a}{\nu_k}, \Delta t)] ^2 / 2 \sigma^2_k}}).
\end{equation}

\noindent Thus, the log-likelihood function becomes
\begin{equation}
    \mathcal{L} =  C - \sum_k \frac{\mid S_k - S_0 W(t_k + t_0 + \frac{a}{\nu_k}, \Delta t) \mid^2}{2 \sigma^2_k}.
\end{equation}

We derived the posterior probability distributions using the nested sampling Monte Carlo algorithm implemented in the \texttt{UltraNest} package$^{\textcolor{blue}{4}}$ \citepads{2021JOSS....6.3001B}. \texttt{UltraNest} combines several advanced methods for efficient nested sampling, including the \texttt{mlfriends} algorithm \citepads{2016S&C....26..383B, 2019PASP..131j8005B}, which adopts ellipsoidal sampling regions based on the local geometry of the likelihood distribution. It also incorporates strategies for dynamic live point management, robust integration with bootstrapped uncertainties, and adaptive termination criteria to ensure accurate results. Uniform priors are adopted for all parameters except for $S_0$, which is assigned a logarithmic prior, meaning log($S_0$) is uniformly distributed. Bursts B20 and B21 show a significant drift. 

The characteristics of these two drifting bursts are shown in Table~\ref{tab:bursttable}. Errors are given by the 95 per cent confidence interval on the posterior distribution. 

\begin{table*}
\caption{Characteristics of the two drifting Stokes~V stellar bursts.}
\label{tab:bursttable}
\renewcommand{\arraystretch}{1.5}
\begin{tabular*}{\textwidth}{@{\extracolsep{\fill}} l l l l l l}
\hline \hline
Burst ID            & Time of arrival ($t_0$)                & Burst flux density ($S_0$)    & Burst width ($\Delta t$)$^{\text{a}}$  & Burst drift delay ($a$)             &  Burst drift rate ($a$)            \\ 
                    & (MJD)                                  & (mJy)                 & (sec)                     & (min)                         &  (MHz sec$^{-1}$)             \\\hline
B20$^{\text{b}}$    & 57525.924011$^{+0.000001}_{-0.000001}$ & -71.4$^{+1.7}_{-1.7}$ & 62.3$^{+0.1}_{-0.1}$      & 0.998$^{+0.004}_{-0.004}$     &  -0.801$^{+0.003}_{-0.003}$    \\  
B21                 & 58552.202540$^{+0.000111}_{-0.000178}$ & -4.5$^{+1.4}_{-1.3}$  & 63$^{+31}_{-11}$          & 13.3$^{+0.4}_{-0.5}$    &  -0.060$^{+0.002}_{-0.002}$     \\ 
\hline \hline
\multicolumn{6}{l}{$^{\text{a}}$ Width at a fixed frequency channel, or the sweep-corrected burst width.} \\
\end{tabular*}
\end{table*}

\subsection{Drifting burst properties}
\begin{figure*}
    \centering
        \includegraphics[width=0.32\textwidth]{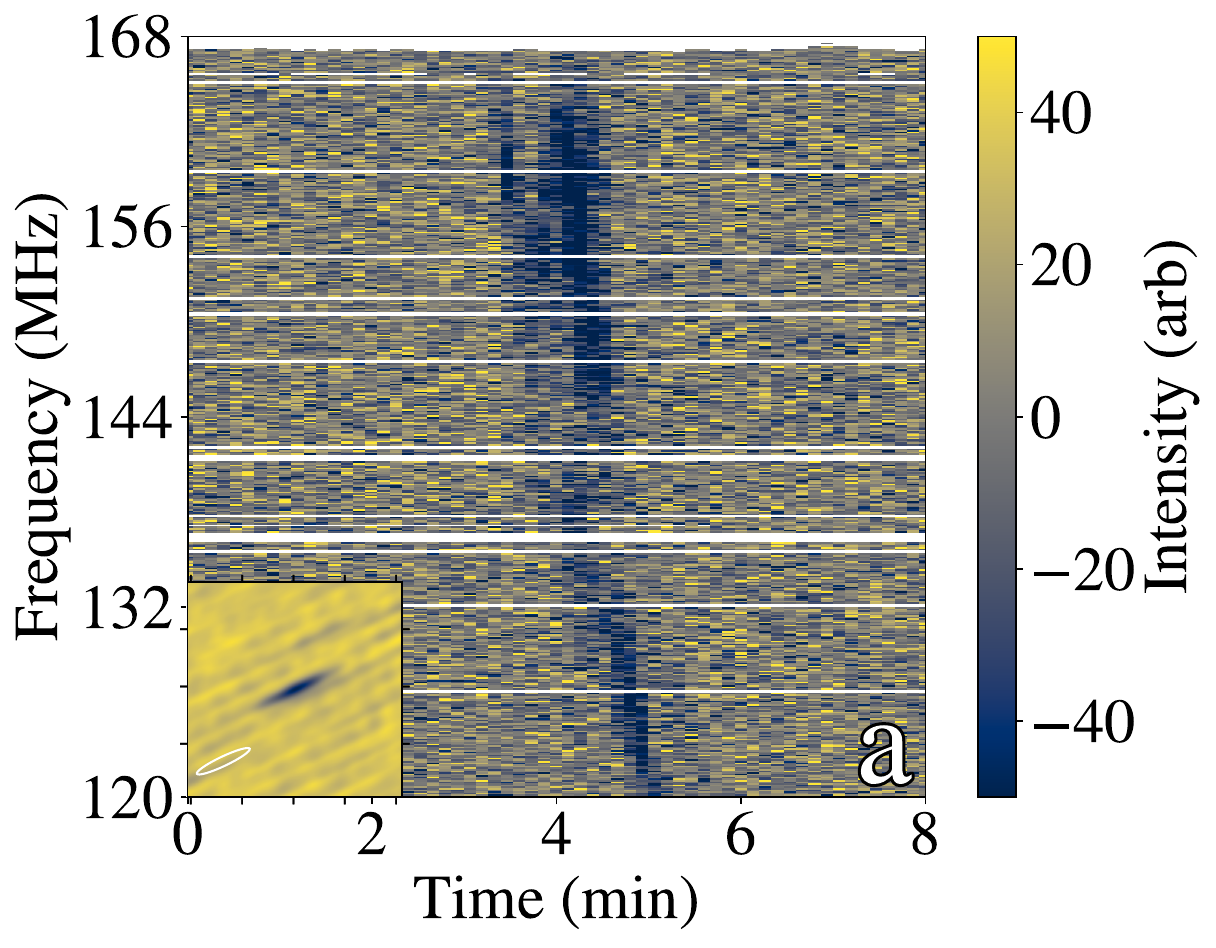}
    \hfill 
        \includegraphics[width=0.32\textwidth]{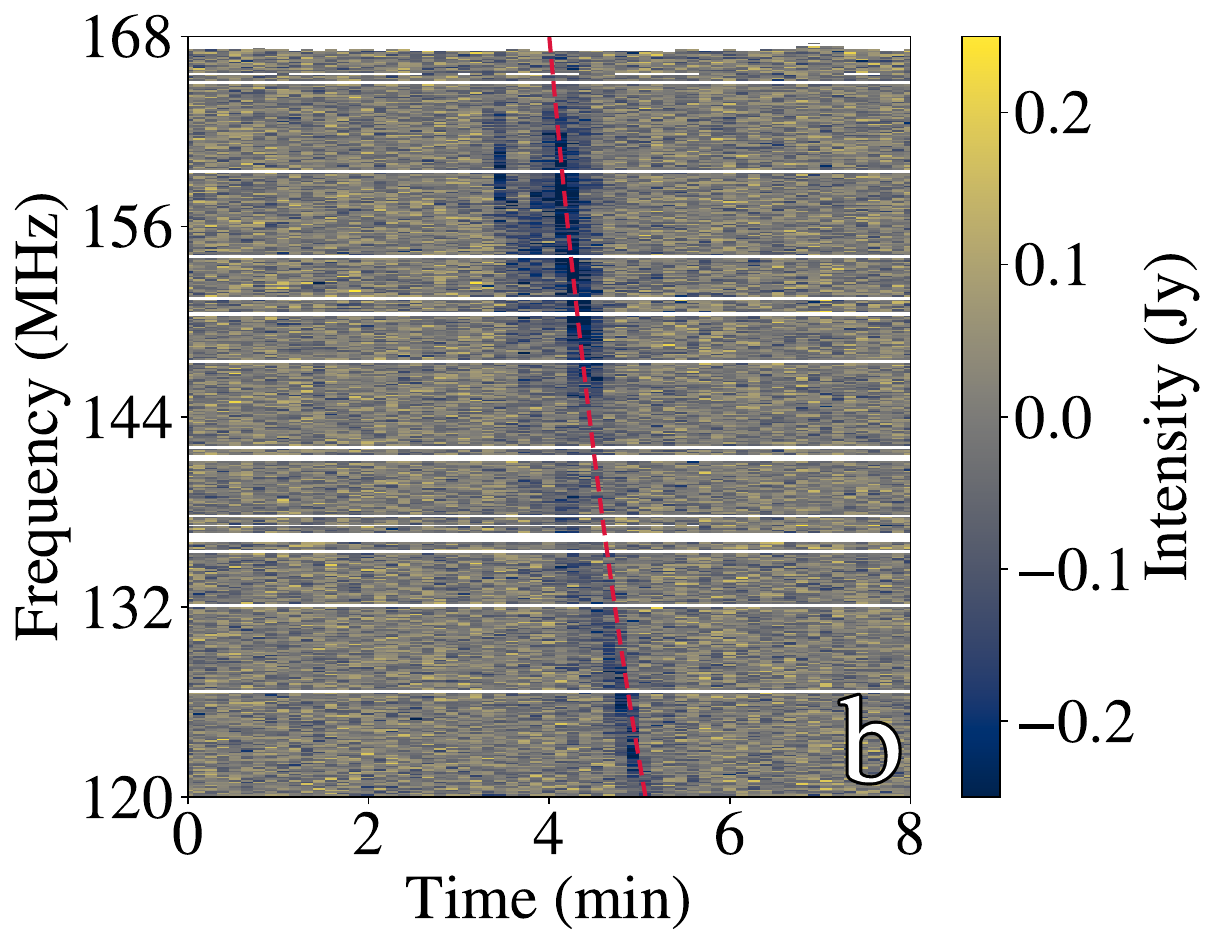}
    \hfill
        \includegraphics[width=0.32\textwidth]{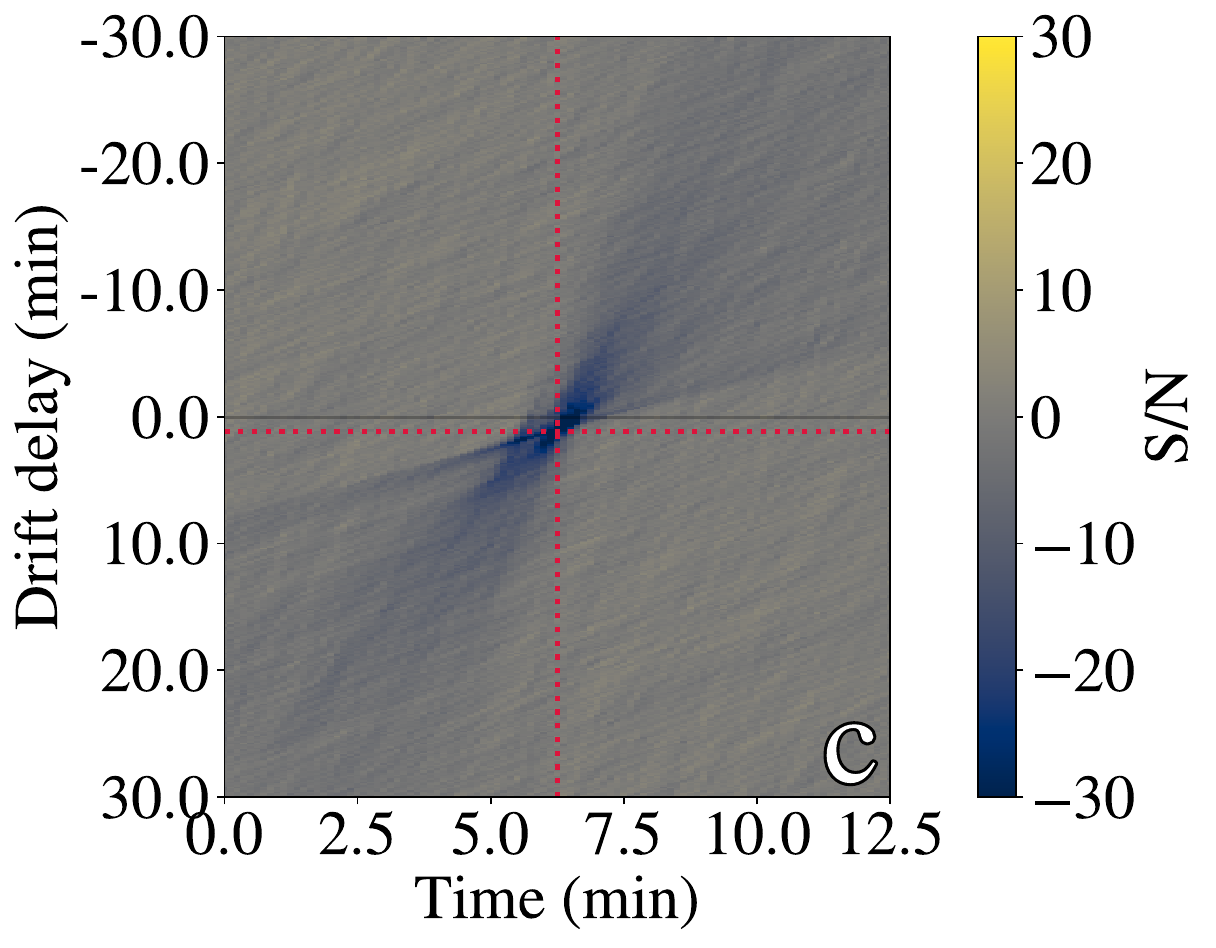}
    \caption{Panel~a: Dynamic spectrum of B20 normalised by the blank-sky variances. The inset in the bottom left displays an interferometric image indicating that the emission originates from a point source. The horizontal white bands are frequency channels masked due to the presence of RFI. Panel~b: Dynamic spectrum of B20, with the highest likelihood drift rate (i.e. the overall drift rate) shown as a dashed red line. Panel~c: Frequency-integrated burst brightness as a function of the modelled drift delay. The intersection of the dotted red lines indicates the location of the highest S/N at $\sim$1\,minute; the horizontal line represents zero drift and corresponds to the original time series.}
    \label{fig:joesburst}
\end{figure*}

\begin{figure*}
    \centering
        \includegraphics[width=0.32\textwidth]{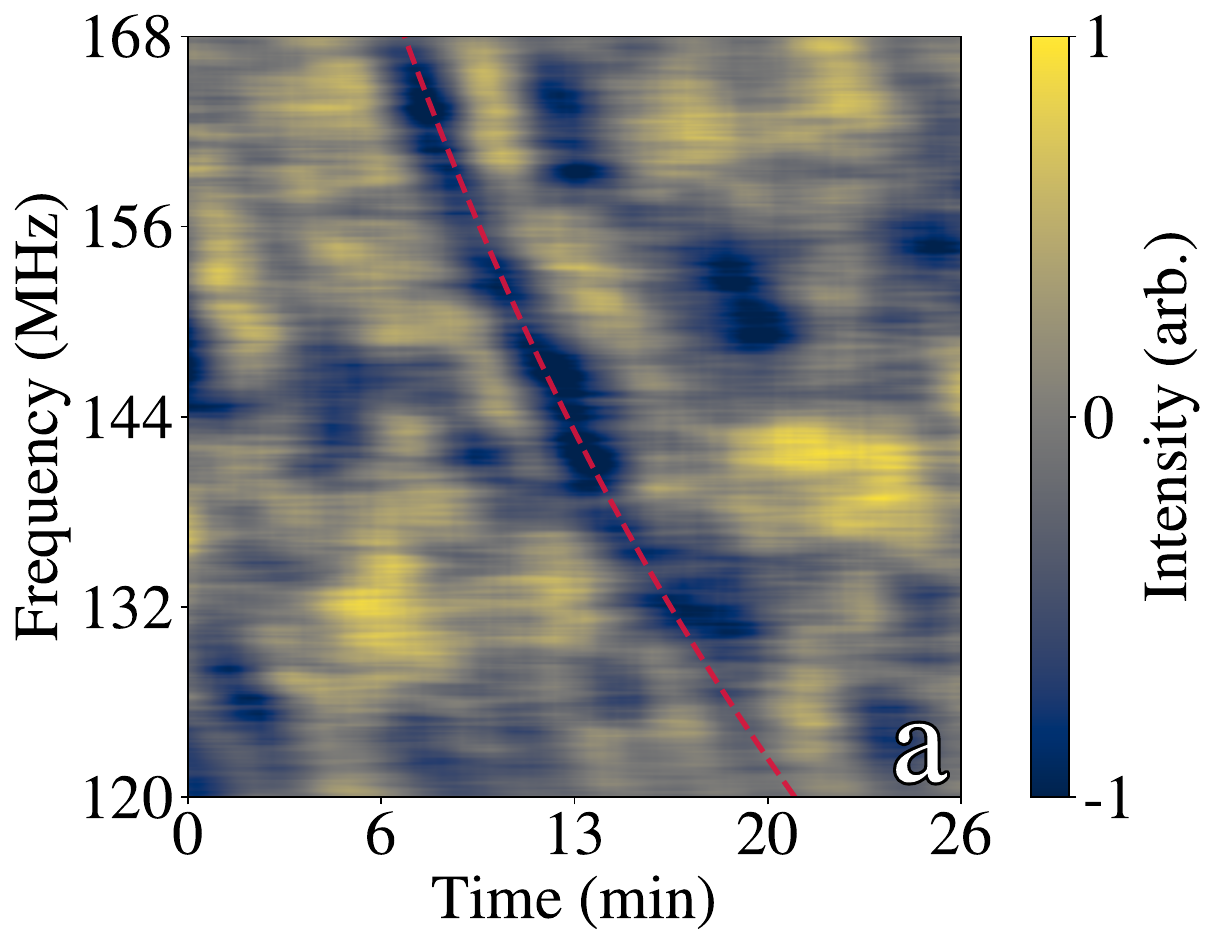}
    \hfill
        \includegraphics[width=0.32\textwidth]{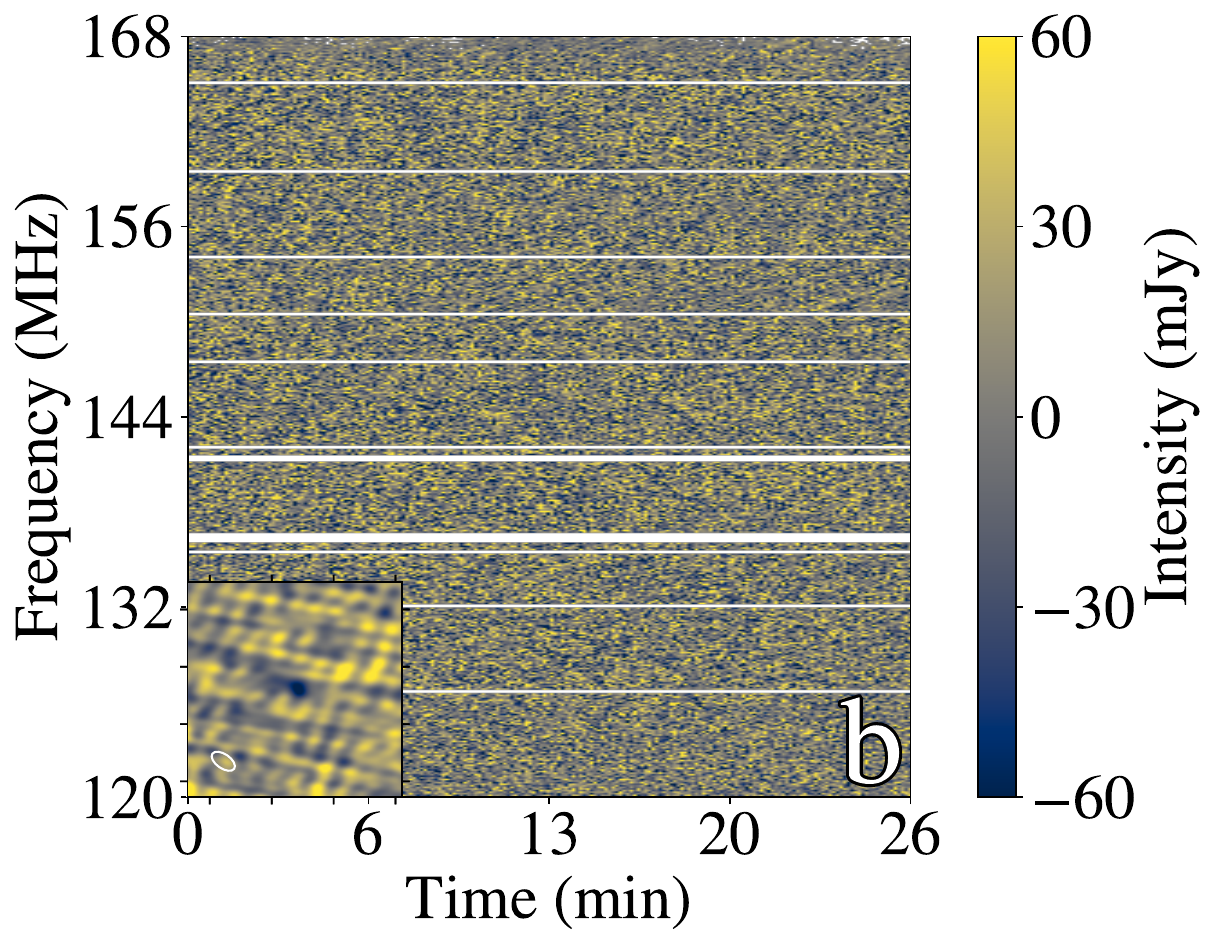}
    \hfill
        \includegraphics[width=0.32\textwidth]{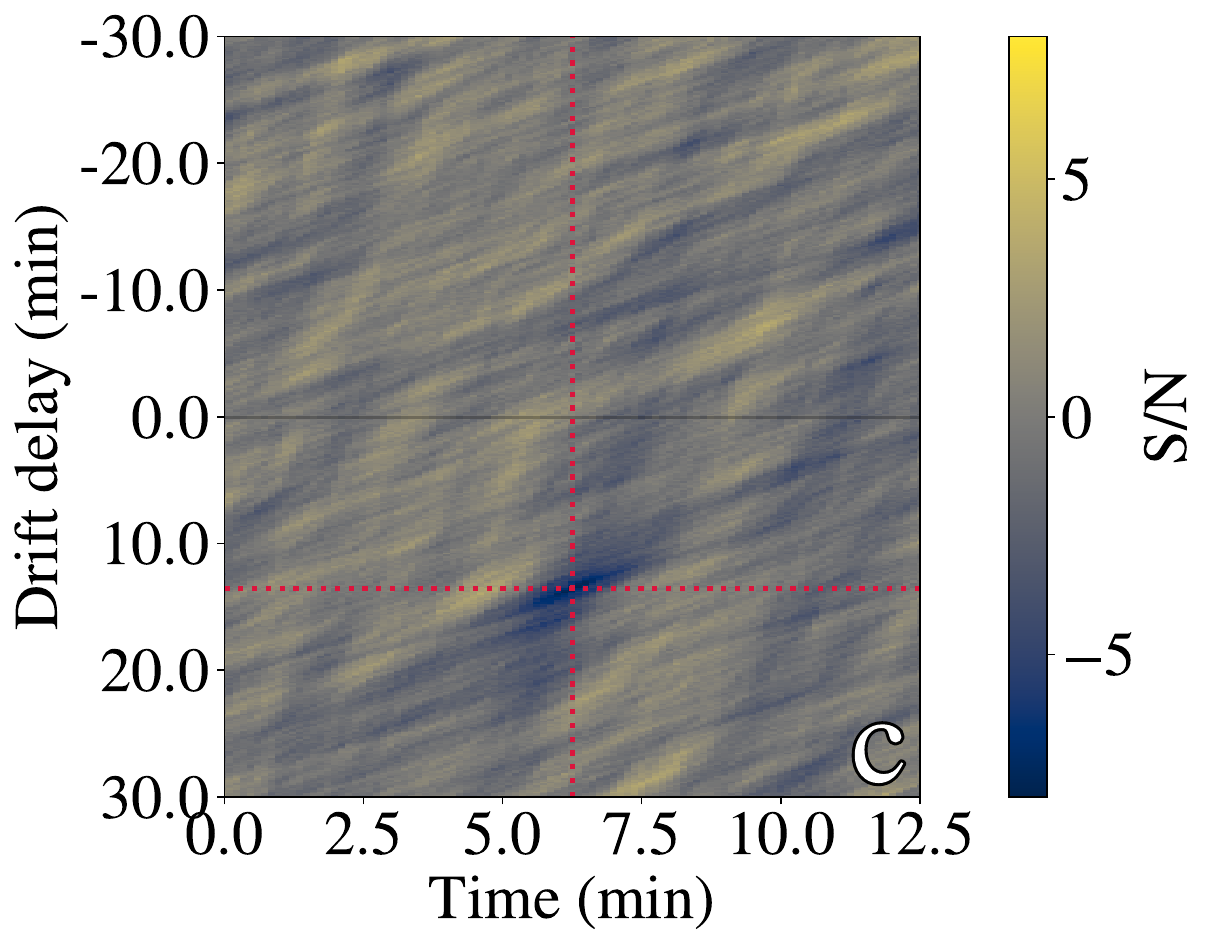}
    \caption{Panel~a: Dynamic spectrum of B21 convolved with a kernel designed to follow the burst’s drift pattern. The drift of the burst is overplotted as a dotted red line. Panel~b: Raw dynamic spectrum of B21. The horizontal white bands are frequency channels masked due to the presence of RFI. The inset in the bottom left shows an interferometric image reconstructed along the dotted line in panel~a, indicating that the emission originates from a point source. Panel~c: Frequency-integrated burst brightness as a function of the modelled drift delay. The dotted cross indicates the location of the highest S/N at a drift of 13 minutes, where the burst aligns vertically and thus appears offset from the horizontal zero drift line, which corresponds to the original time series.}
    \label{fig:13minburst}
\end{figure*}

Burst B20, shown in Fig.~\ref{fig:joesburst}, and burst B21, shown in Fig.~\ref{fig:13minburst}, both show a time-frequency drift rate: -0.801$^{-0.003}_{+0.003}$ and -0.06$^{-0.002}_{+0.002}$\,MHz\,sec$^{-1}$, respectively. We did not analyse burst B20 in detail as it has been covered by \citet{Callinghamsubmitted}. 

\addtocounter{footnote}{4}
\footnotetext[4]{\url{https://johannesbuchner.github.io/UltraNest/}}

The modelled burst flux for B21, integrated over the entire burst, is $4.5^{+1.4}_{-1.3}$\,mJy in Stokes~V, which places it at the edge of our detection sensitivity limit. The burst originates from stellar object LP\,215-56, (Gaia DR3 ID \texttt{769964666964655488}), an M2.4V dwarf star \citepads{2020ApJ...892...31B}, located at a distance of 62.13\,pc \citepads{2020yCat.1350....0G}. The burst is significantly circularly polarised, with our measurements indicating a polarisation fraction of greater than 50 per cent. The presence of a nearby ($<18'$) 1.3\,Jy radio source imposes dynamic range limitations in the Stokes I data, preventing us from determining a more precise polarisation fraction beyond this lower limit of 50 per cent. The star is likely part of a binary system, as its \textit{Gaia} re-normalised unit weight error is 5.25; values larger than 1.4 indicate binary systems \citep{lindegren2018re}. 

The star has been observed at optical wavelengths by the Transiting Exoplanet Survey Satellite (TESS). We searched the TESS 2-minute light curve data (\texttt{TIC 253052423}; sectors 22, 48, and 49) and find no flares, which allows us to place an upper limit on the flare rate of $<1$ flare yr$^{-1}$ for flare energies $\geq 10^{33}$\,erg, assuming a typical power law slope for the flare frequency distribution of $-2$~\citep{2021A&A...645A..42I}. The upper limit energy was determined by defining the smallest detectable flare as three consecutive data points $3\sigma$ above the noise level of the de-trended light curve. From the so derived minimum detectable equivalent duration\citepads{1972Ap&SS..19...75G}, we derived a bolometric energy following the procedure in\citeads{2013ApJS..209....5S}, and assuming a blackbody temperature of $10,000$\,K. This gives a detection threshold of $3\times10^{32}$\,erg in a $77\,$d long light curve. The Hungarian-made Automated Telescope Network (HATNet; \citeads{2011AJ....141..166H}) Exoplanet Survey reports a rotational period of $\sim$44\,days \citepads{2018haex.bookE.111B}. However, we are unable to confirm this with the TESS light curve data due to the limited observation time. 

As the binary status of the star is uncertain, we considered two scenarios: (i) the star is not in a binary system and (ii) the more likely scenario is that the star is in a binary system. If the star is not in a binary system, we estimated its mass and radius based on its position in the \textit{Gaia} colour-magnitude diagram: $M = 0.507 \pm 0.020 M_\odot$ and $R = 0.509 \pm 0.015 R_\odot$ \citepads{2019yCat.4038....0S}. Alternatively, if the star is in a binary system, we made the rough assumption that our target star has half the mass $M = 0.25 M_\odot$ and an accompanying radius of $R = 0.40 R_\odot$. Using these mass and radius values, we could estimate the brightness temperature of the emission.

The brightness temperature $T_{B}$ of the emission is given by
\begin{equation}
    T_B = 10^{-26} \frac{c^2}{2 k_B \nu^2} \frac{S}{\Omega} \hspace{0.25cm} \text{K} = 10^{-26} \frac{c^2}{2 k_B \nu^2} \frac{S d^2}{\pi R^2} \hspace{0.25cm} \text{K},
\end{equation}
\begin{figure*}
  \begin{minipage}[c]{0.67\textwidth}
    \includegraphics[width=\textwidth]{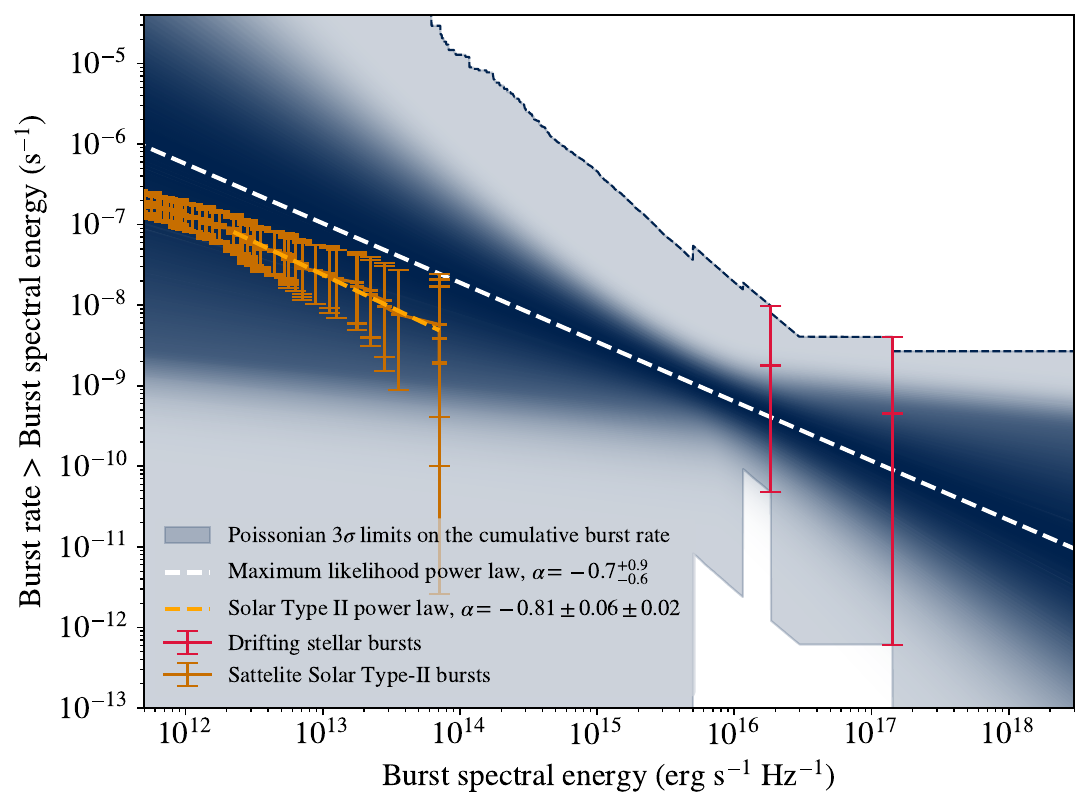}
  \end{minipage}\hfill
  \begin{minipage}[c]{0.315\textwidth}
    \caption{
       Burst rate of circularly polarised drifting stellar radio bursts that exceed a given spectral energy threshold (in erg\,s$^{-1}$\,Hz$^{-1}$). The monochromatic grey area represents the region spanned by the 3$\sigma$ Poissonian limits; the dark dashed line indicates the upper limit. The two red data points represent detected drifting bursts, which cause sudden jumps in the Poisson lower limit when specific energies reach the detection threshold for that star. For energies exceeding that of a drifting burst, the lower limit drops again due to insufficient data on higher-energy bursts. The dashed white line represents the maximum likelihood of the cumulative luminosity distribution, while the blue diverging shaded region shows other possible solutions, with faded colours indicating a lower likelihood. The cumulative spectral energy distribution of decametric solar Type~II radio bursts, to which a power law has been fitted using the Python \texttt{powerlaw} package \citepads{2009SIAMR..51..661C}, is plotted in orange.
    } \label{fig:eventrates_comp}
  \end{minipage}

\end{figure*}
\noindent where $c$ is the speed of light in cm/s, $k_B$ the Boltzmann constant in erg/K, $\nu$ the frequency of the emission in Hz, $S$ the flux density of the burst in mJy, $\Omega$ the source solid angle in sr, $d$ the distance to the source in cm, and $R$ the radius of the emitting region in cm. To obtain the second equality, we assumed that the emission region has a projected size comparable to that of the stellar disk. For our $4.5^{+1.4}_{-1.3}$\,mJy burst, assuming that the star is not in a binary system, the brightness temperature is $6.1 \times 10^{13}$\,K. If the star is in a binary system with $R = 0.40 R_\odot$, the brightness temperature increases to $9.9 \times 10^{13}$\,K. Such high brightness temperatures at this frequency require a coherent radio emission process, which can only be produced by either plasma emission or through the electron cyclotron maser instability (ECMI;\citeads{2006A&ARv..13..229T}).

The linear polarisation fraction of solar Type~II bursts coinciding with CMEs is minimal, often approaching zero (\citealt{1973SoPh...29..149G,2022SoPh..297...47M,2024EGUGA..2614421D}). However, burst B20 does show weak linear polarisation features, as described by \citet{Callinghamsubmitted}. For burst B21, we calculated the Faraday rotation measure and confirmed that our burst does not emit any significant linear polarisation components (see Appendix~\ref{appendix:faraday}).

\subsection{Type~II burst rate}

With two detections of drifting bursts that exhibit characteristics consistent with Type~II bursts, including their drift rates, emission frequencies, durations, and brightness, we can establish constraints on the Type~II-like burst rate using two separate but complementary methods. The first is an estimate of the Type~II burst rate based on observed detections, which constrains the burst rate by considering only the effective sample of stars from which bursts could be detected. This expands on the method presented by \citet{Callinghamsubmitted}, applying it to each star rather than a single source. With the second approach, we parametrised the cumulative Type~II-like burst luminosity distribution, $f(L),$ as a power law and calculated for which parameters the likelihood is maximised. Both methods are visualised in Fig.~\ref{fig:eventrates_comp} and are detailed below.

% \begin{figure*}
%     \centering
%     \includegraphics[width=1\linewidth]{Figures/TypeII_eventrates_with_solartype2.png}
%     \caption{The burst rate of circularly polarised drifting stellar radio bursts that exceed a given spectral energy threshold (erg\,s$^{-1}$\,Hz$^{-1}$). The monochromatic grey region represents the region spanned by the 3$\sigma$ Poissonian limits, where the dark dashed line indicates the upper limit. The two red data points with error bars represent detected drifting bursts, causing sudden jumps in the Poisson lower limit when specific energies reach the detection threshold for that star. For energies exceeding a drifting burst, the lower limit drops again due to insufficient data on higher-energy bursts. The white dashed line represents the maximum likelihood of the cumulative luminosity distribution, while the blue diverging shaded region shows other possible solutions, with fading colour indicating a lower likelihood. The cumulative spectral energy distribution of decametric Solar Type~II radio bursts is plotted in orange, through which a power law has been fitted using the Python \texttt{powerlaw} package \citepads{2009SIAMR..51..661C}.}
%     \label{fig:eventrates_comp}
% \end{figure*}

For our more observational approach, we calculated the burst rate for drifting Type~II-like bursts with spectral energies (erg\,s$^{-1}$\,Hz$^{-1}$) exceeding a given energy threshold ($E$) as 

\begin{equation} 
R_{\text{cum}}(E) = \frac{N_\text{eff}(\geq E)}{T_{\text{eff}}(E)}, 
\end{equation}

\noindent where $N_{\text{eff}}$ is the effective cumulative amount of detected bursts with $\geq E$ from the effective sample of stars from which a burst with energy $E$ could be detected, given our sensitivity of $\sim$2.5\,mJy for bursts of 1\,minute. $T_{\text{eff}}$ represents the total time spent observing this effective star sample. This estimation is slightly conservative as it does not include extremely bright bursts (e.g. $\geq 10^{18}$\,erg\,s$^{-1}$\,Hz$^{-1}$) from stars past 100\,pc. 

We show the Poissonian 3$\sigma$ upper and lower limits for the burst rate per day as the grey monochromatic region in Fig.~\ref{fig:eventrates_comp}. Our two detections of drifting bursts significantly increase the lower limit of the burst rate, causing pronounced jumps when the energy surpasses the threshold at which the host stars of the drifting bursts can be observed. The first jump at lower energies is caused by burst B20, whereas the second higher jump is caused by both B20 and B21. The true luminosity distribution of these drifting radio bursts, regardless of its exact form, should lie within this grey monochromatic region. The approach by \citet{Callinghamsubmitted} can be used for bursts with radio luminosities similar to that of burst B20, whose Type~II burst rate they determined to be $0.84^{+1.10}_{-0.14} \times 10^{-3}$ per day. This is consistent with our upper limit for bursts with spectral energies of $\sim$$10^{16}$\,erg\,s$^{-1}$\,Hz$^{-1}$ or higher. 

\begin{figure}
    \centering
    \includegraphics[width=1\linewidth]{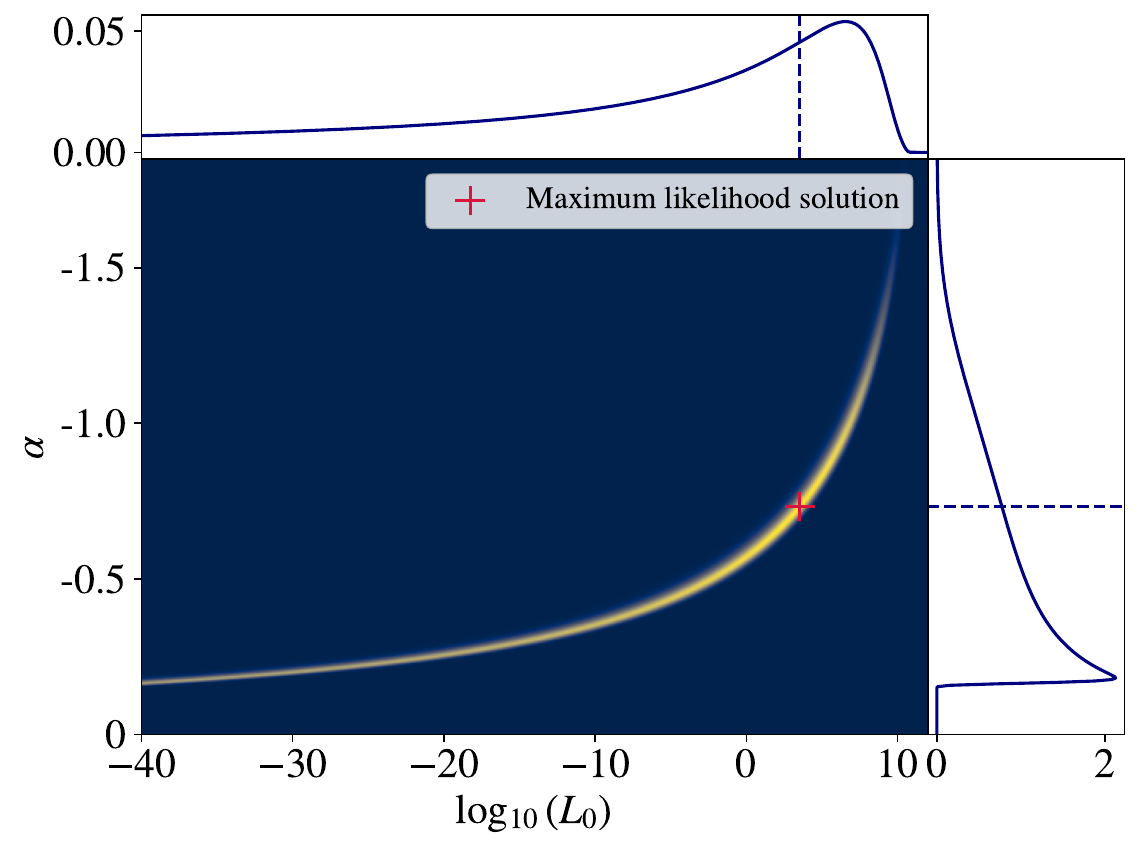}
    \caption{Joint posterior probability distribution for the cumulative luminosity distribution parameters $\alpha$ and $L_0$. The maximum likelihood is indicated with a plus-sign, and is reached when $\alpha=-0.7$, and $L_0 = 10^{3.5}$\,erg\,s$^{-1}$\,Hz$^{-1}$. The marginalised PDFs are shown on the top and right axes, respectively.}
    \label{fig:joint_pdf}
\end{figure}

In our second approach, we parametrised the cumulative burst luminosity distribution, $f(L),$ as a power law:

\begin{equation}
    f(L) = \left(\frac{L}{L_0}\right)^\alpha ,
\end{equation}

\noindent where $f(L)$ is the number of bursts with luminosity $L$ exceeding some value per unit time, $L_0$ is the normalisation point where we expect on average one burst per unit time, and $\alpha$ is the power law index. We can think of each observation of a star as a single experiment ($i$), where we have a threshold luminosity $L_i^{\text{th}} = 4 \pi d_i^2 S_i^{\text{th}}$ for detection. The total likelihood across all observations of all stars is given by 

\begin{equation}
    \mathcal{L}(L_0, \alpha) = \prod_i \frac{\text{exp}(-\lambda_i)\lambda_i^{k_i}}{k_i!},
\end{equation}

\noindent where

\begin{equation}
    \lambda_i = t_i \left( \frac{4 \pi d_i^2 S_i^{\text{th}}}{L_0} \right)^\alpha, 
\end{equation}

\noindent where $k_i$ is the number of detected drifting bursts from experiment $i$. We calculated the likelihood across a range of values, which yielded the two-dimensional posterior distribution for $\alpha$ and $L_0$, shown in Fig.~\ref{fig:joint_pdf}. For the priors, we used a logarithmic prior for $L_0$, meaning that log($L_0$) is uniformly distributed, while $\alpha$ is uniformly distributed. The main panel shows the joint posterior distribution, whereas the right and top panel shows the marginalised probability density functions (PDFs) for $\alpha$ and $L_0$, respectively. The maximum likelihood solution is attained at $\alpha =-0.732$, and $L_0 = 10^{3.528}$\,erg\,s$^{-1}$\,Hz$^{-1}$, and is highlighted in Fig.~\ref{fig:eventrates_comp} as a white dashed line. For $\alpha$ we calculated the 99.7 per cent confidence interval as $\alpha=-0.7^{+0.9}_{-0.6}$, $L_0$ is weakly constrained. Combinations of $\alpha$ and $L_0$ with a likelihood above 0.01 collectively form the diverging shaded region in the figure, where the intensity of the shading reflects the relative likelihood of each solution, with darker regions corresponding to higher likelihoods. As the likelihood decreases, the shading becomes more transparent, effectively conveying the gradual transition from the most likely solutions to less probable ones. 
\begin{figure}
    \centering
    \includegraphics[width=1\linewidth]{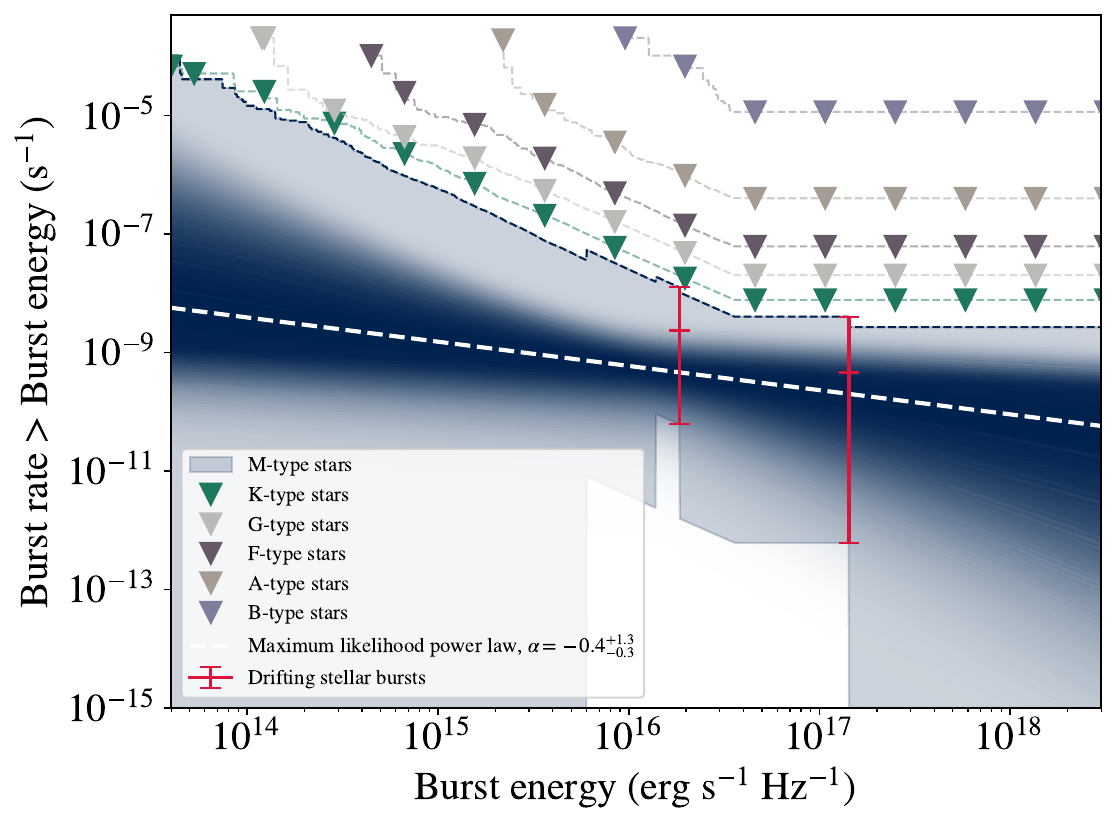}
    \caption{Burst rate of circularly polarised drifting stellar radio bursts that exceed a given spectral energy threshold (erg\,s$^{-1}$\,Hz$^{-1}$) based on the spectral type of the star observed. As in Fig.~\ref{fig:eventrates_comp}, we plot a grey monochromatic region and diverging shaded region, but now only for M~dwarfs, as two drifting bursts have been detected from this spectral type. The dashed white line again represents the maximum likelihood of the cumulative luminosity distribution only fitted to M~dwarfs, while the blue diverging shaded region shows other possible solutions, with fading colour indicating a lower likelihood. The Poissonian upper limits for other spectral types observed are shown as dashed lines, with triangle markers emphasising their status as upper limits.}
    \label{fig:eventrates_spec}
\end{figure}
Both methods yield consistent estimates for the burst rate, with the more observational approach providing robust bounds and the parametric distribution offering insights into the underlying luminosity distribution. Additionally, we plot the cumulative spectral energy distribution of decametric solar Type~II bursts from \citetads{2024A&A...691L...8M} in Fig.~\ref{fig:eventrates_comp}. The solar Type~II data were observed by the Radio and Plasma Wave Investigation (WAVES) instruments on board the Wind, STEREO A, and STEREO B spacecrafts. We assumed that the number of Type~II events collated by\citetads{2024A&A...691L...8M} corresponds to all Type~IIs observed during the quoted period from November 2006 to July 2023. We assumed Poissonian errors on the burst numbers and a fractional error of 20 per cent for the reported peak flux values. We fit a power law to the distribution using the Python \texttt{powerlaw} package\citepads{2009SIAMR..51..661C}. In addition to performing power law fits, this package identifies the initial turnover point in the cumulative distribution to determine the appropriate minimum spectral energy from which to start the fit. The fitted power law has a slope of $-0.81 \pm 0.06 \pm 0.02$, where the first error is determined as the quadrature sum of the error on the fit using the fractional 20 per cent peak flux errors and the second error is determined by bootstrapping.

\begin{figure*}
    \centering
        \includegraphics[width=0.32\textwidth]{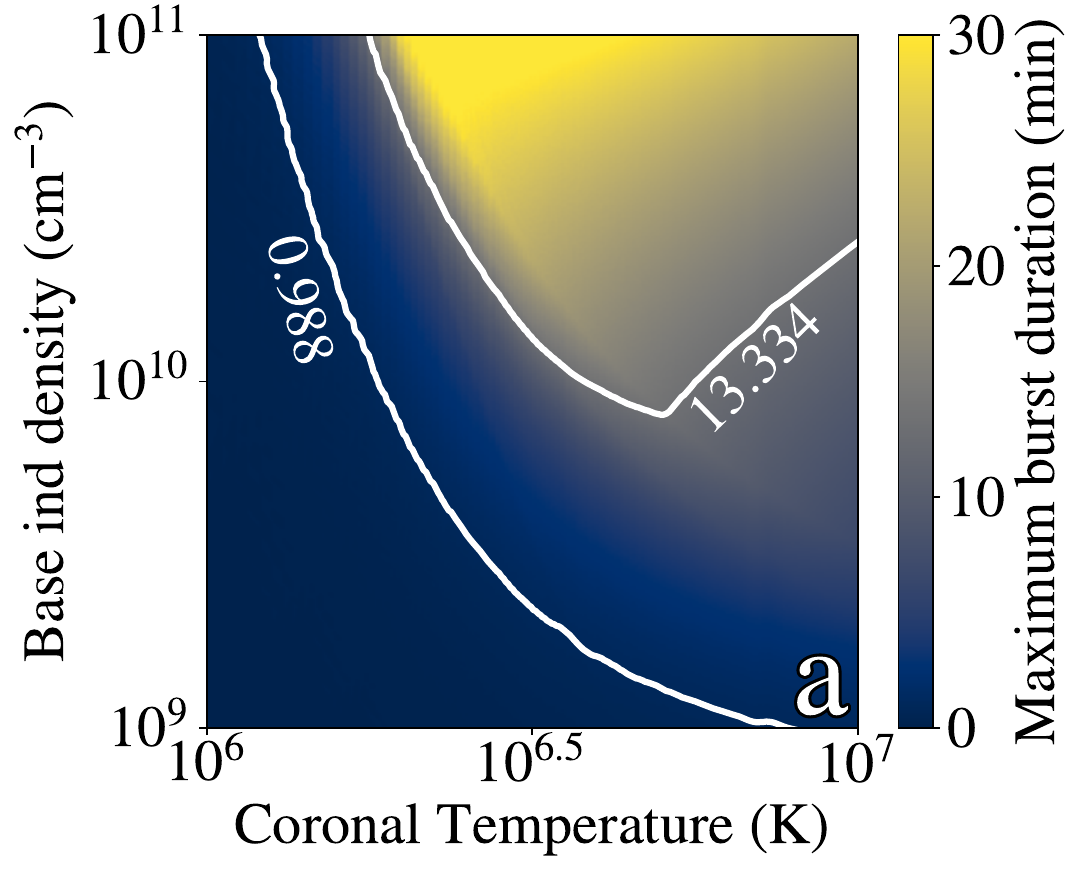}
    \hfill
        \includegraphics[width=0.32\textwidth]{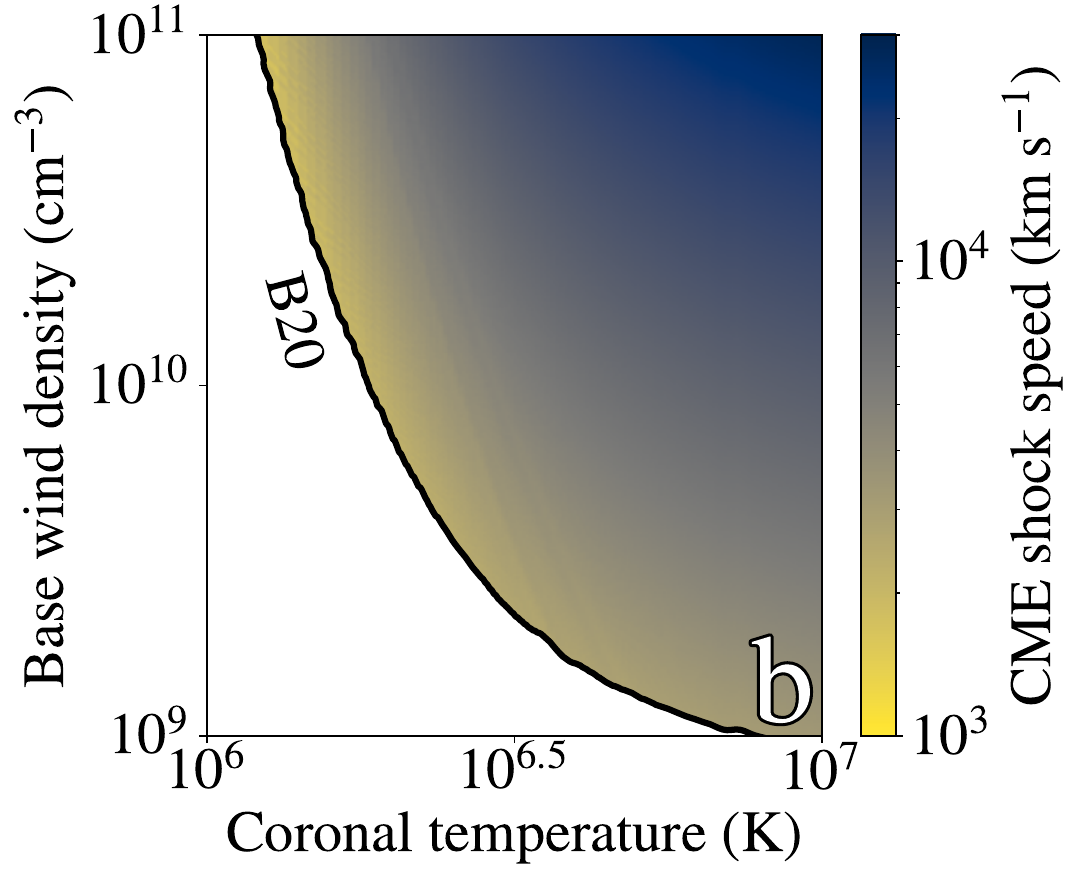}
    \hfill
        \includegraphics[width=0.32\textwidth]{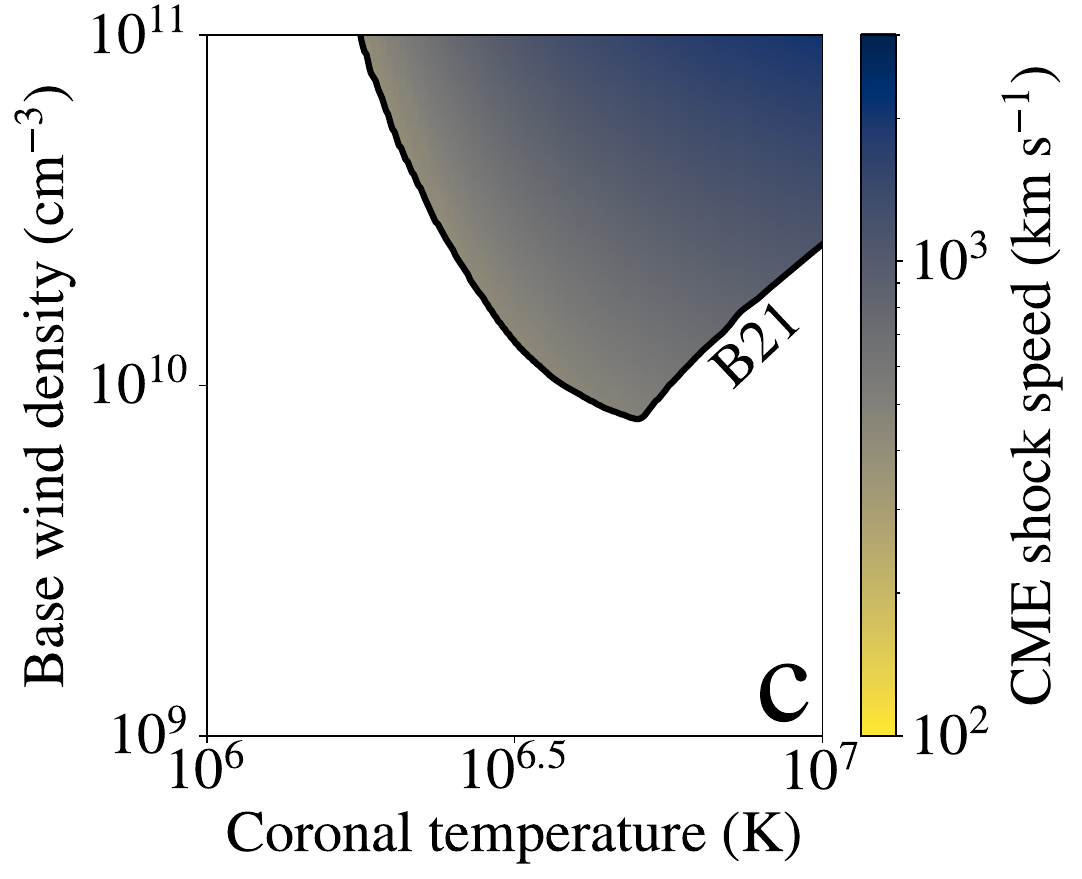}
    \caption{Panel~a: Maximum burst duration in minutes through our observing band (120-168\,MHz) as a function of coronal temperature and base wind density, modelled with an iso-thermal Parker wind model. White contour lines show the observed durations of bursts B20 and B21. Regions above the contours can produce bursts with the observed durations, with CME shock speeds exceeding the wind- and Alfv\'{e}n speed. Panel~b: CME shock speeds that exactly produce burst B20's observed duration of 0.988\,minutes, as a function of coronal temperature and base wind density, modelled with an isothermal Parker wind model. The black line marks the minimum shock speed exceeding both the Alfv\'{e}n and wind speeds. Solutions exist for shock speeds upwards of $\sim$1000\,km\,s$^{-1}$. Panel~c: CME shock speeds that exactly produce burst B21's observed duration of 13.334\,minutes, as a function of coronal temperature and base wind density, modelled with an isothermal Parker wind model. The black line marks the minimum shock speed exceeding both the Alfv\'{e}n and wind speeds. Solutions exist for shock speeds upwards of $\sim$300\,km\,s$^{-1}$.}
    \label{fig:shockmodel}
\end{figure*}

By restricting the total observation time to stars of a specific spectral type, we can refine our constraints based on the spectral type of the star. This is shown in Fig.~\ref{fig:eventrates_spec}. We assigned stars to spectral types based on their \textit{Gaia} colours, using an approximate binning that is sufficient for our analysis. Stars were classified as M, K, G, F, A, or B types if their colours fell within the ranges 4.25-2.27, 2.27-1.08, 1.08-0.71, 0.71-0.33, 0.33 to -0.04, and -0.04 to -0.49, respectively. Using the more observational approach, we plot the upper limits for stars of spectral types for which no drifting bursts were detected. For M~dwarfs, where we have two detections of drifting bursts, we again also parametrised and fitted the cumulative luminosity distribution, $f(L),$ to estimate the burst rate. The maximum likelihood is achieved for $\alpha = -0.4$, and $L_0 = 10^{2.1}$\,erg\,s$^{-1}$\,Hz$^{-1}$.

\section{Discussion}  \label{sec:disc}
In our $\sim$100\,stellar-year search, we detected 21\,stellar radio bursts, 20\, of which originated from M~dwarfs. Given that M~dwarfs are the most common star type, and exhibit high stellar activity and high magnetic field strengths, it is unsurprising that the majority of radio bursts were found in these stars. Only B12 originated from a RSCVn close binary. Among all 21\,bursts, only B20 and B21 exhibit any measurable form of drift.

B20 and B21 are drifting coherent radio bursts, which can be produced either by plasma emission, where the burst would likely be tracing a CME, or by ECMI emission. 
Stellar ECMI bursts have been reported previously (e.g.\citeads{2008ApJ...674.1078O}, \citeads{2008ApJ...684..644H}, \citeads{2025A&A...695A..95Z}), but they are typically different with respect to their durations and morphology, and are often modulated periodically with the host’s rotation.
Assuming our two drifting bursts are produced by plasma emission, based on their specific drift structure, burst B21 could trace the backbone of a Type~II burst, while the remainder of the flux of the burst has fallen below our sensitivity threshold. Both fundamental and harmonic emission could produce a burst with brightness temperatures of $\sim$10$^{14}$\,K \citepads{2001A&A...374.1072S}. However, harmonic emission cannot reach polarisation fractions $>$50 per cent \citepads{2021MNRAS.500.3898V}. Burst B21 is significantly circularly polarised, 50-100 per cent, and therefore if the burst is produced by plasma emission, it is likely fundamental emission. Burst B20 is likely produced by plasma emission, as argued by \citet{Callinghamsubmitted}.

Assuming plasma emission, we calculated the corresponding shock speed of a burst based on its drift rate, enabling a direct comparison with solar CMEs. The shock speed must exceed both the stellar wind speed and the Alfv\'{e}n speed for the shock to exist, and is therefore dependent on the physical properties of the stellar environment. We modelled the stellar wind using an isothermal Parker wind model with base density and temperature as parameters. The Parker wind model is used as a first-order approximation, since the complex M-dwarf wind structures are not well constrained \citepads{1999isw..book.....L}, and more sophisticated models would not qualitatively change our first-order comparison of CME speeds. We assumed the magnetic field to fall off as radial distance to the power of $-3$ (i.e. dipolar) out to a distance where the magnetic energy density equals the wind kinetic energy density. Beyond this radial distance we assumed that the field lines open up and fall of as distance to the power of $-2$. With the wind density and field specified, we can calculate the Alfv\'{e}n speed at any radial distance. For both the host stars of bursts B20 and B21, we assumed a mass of 0.5\,M$_\odot$ and a radius of 0.5\,R$_\odot$, as variations in these parameters minimally affect the results. Additionally, we assumed a surface magnetic field of 100\,Gauss. With various base wind densities and coronal temperatures, we determined the minimum radial speed of a shock required to exceed both the wind speed and the Alfv\'{e}n speed. Next, by transforming the radial density profile to the emission plasma frequencies, we could calculate the maximum duration of a burst that could pass through our 120-168\,MHz observing band. This is shown in Fig.~\ref{fig:shockmodel}a, where the white contour lines represent the durations of the drifting bursts B20 and B21. Solutions with shorter burst durations are also allowed as the CME shock speed can exceed the minimum speed required to surpass the wind and Alfv\'{e}n speeds. The shock velocity ranges that could exactly produce burst B20 and burst B21 are shown in Figs.~\ref{fig:shockmodel}b and~\ref{fig:shockmodel}c, respectively. This demonstrates that a shockwave within an isothermal Parker wind model can produce the emission as seen in burst B20 with shock speeds upwards of $\sim$1000\,km\,s$^{-1}$, which agrees with the shock speed of 2400\,km\,s$^{-1}$ quoted by \citet{Callinghamsubmitted}. For burst B21, the emission can be produced by shock speeds ranging from $\sim$300\,km\,s$^{-1}$ upwards to a few thousand km\,s$^{-1}$, where the lower end is more likely as lower wind density scenarios are preferred. These speeds broadly agree with solar CME speeds, although burst B21 is notably on the slower side.

The ECMI can also produce coherent radio bursts, although this requires specific magnetic field configurations, making such scenarios quite rare \citepads{2025A&A...695A..95Z}. The detection of identically drifting harmonic plasma emission would provide convincing evidence in favour of the CME scenario. Furthermore, observing CME signatures across multiple wavelengths would also amount of definitive evidence (e.g.\citeads{2014ApJ...795...68Z}; \citeads{2020ApJ...899...12Y}; \citeads{2024ApJ...961...23N}). However, given the extremely low burst rate (1.0\,year$^{-1}$ with  E\,$>$\,$6.8\times$\,$10^{13}$\,erg\,s$^{-1}$\,Hz$^{-1}$) and the challenges associated with coordinating multi-wavelength observations of the same source, this remains a difficult avenue to pursue. Nevertheless, with two detected drifting bursts, the burst rate presented in Fig.~\ref{fig:eventrates_comp} can be interpreted, at best, as the occurrence rate of stellar CMEs, or, more conservatively, as the occurrence rate of drifting stellar radio bursts. This interpretation is particularly useful for guiding the survey design for future searches for CMEs.

The cumulative spectral energy distribution of solar Type~II bursts has been overplotted, and the power law index from our cumulative luminosity distribution, $\alpha = -0.7^{+0.9}_{-0.6}$, is in agreement with the solar Type~II power law index $\alpha = -0.81 \pm 0.06 \pm 0.02$. Notably, the cumulative spectral energy distribution of solar Type~II bursts lies within our monotonic $3\sigma$ Poisson upper limits and slightly below our maximum likelihood model. This suggests that the low number of detected stellar Type~II bursts is simply be due to a lack of sensitivity and time-on-sky. 

The solar CME-flare relation is often extrapolated to active stars where CMEs might be more readily detectable (e.g.\citeads{2019ApJ...877..105M}). However, this extrapolation remains largely unverified. We explored this by comparing our measured Type~II burst rate to the M~dwarf flare frequency distribution reported by \citet{2019ApJS..241...29Y}. These authors parameterise their distribution as
\begin{equation}
\frac{\text{d}N}{\text{d}E} = C E^{-\alpha},
\end{equation}
where $C$ is a normalisation constant and where they determined $\alpha = 2.13$. We extrapolated this power law and compared it with our observed cumulative Type~II burst rate using the point where the variance on our burst rate is the lowest: a rate of one event per $10^9$\,s, corresponding to energies $\geq 10^{16}$\,erg\,s$^{-1}$\,Hz$^{-1}$.

The cumulative flare rate for M~dwarfs can be written as
\begin{align}
N(\geq E) &= \int_E^{\infty} C E^{-2.13} \, dE \\
&= \frac{C}{1.13} E^{-1.13},
\end{align}
where we estimate $C \approx 7 \times 10^{40}$ from Fig.~3 of \citet{2019ApJS..241...29Y}. Substituting our burst rate into this equation yields a flare energy threshold of $E \approx 2.6 \times 10^{37}$\,ergs. This suggests that the rarity of our observed Type~II radio bursts is consistent with stellar flares having energies $\gtrsim 10^{37}$\,ergs, supporting a potential connection between high-energy flares and detectable radio emission. For context, the most energetic solar flares typically reach $\sim$10$^{32}$\,ergs, meaning that these radio-associated events on M~dwarfs would be orders of magnitude more powerful than the strongest solar flares. This could suggest that only the most energetic flares produce detectable radio emission with LOFAR.

Future observations, for example with the Square Kilometre Array (SKA), offer the potential to directly test the predictions of our maximum likelihood power law. SKA1-Low phase~1 \citep{2022JATIS...8a1014M} is estimated to have roughly an order of magnitude sensitivity increase compared to LOFAR, and an instantaneous field-of-view that is around half of that of LOFAR's High Band Antennas (40\,m versus 30\,m diameter stations). These two taken together imply that in a survey with comparable telescope time-on-sky as ours the expected numbers of burst detection is $2\times 0.5 \times 10^{1.5}=30$. 

The minimum detected burst energy at 100\,pc for SKA1-LOW is $3 \times 10^{15}$\,erg\,s$^{-1}$\,Hz$^{-1}$. At these energies, our derived 3$\sigma$ Poissonian burst rate upper limit is around $5 \times 10^{-8}$\,sec$^{-1}$, while our maximum likelihood power law indicates a burst rate of $1.5 \times 10^{-9}$\,sec$^{-1}$. To estimate the observation time required for SKA to probe these regions, we assumed a stellar sample similar to this study, including all stars within 100\,pc. This is a conservative estimate, as SKA’s greater sensitivity could detect bursts from more distant stars. Given this sample, the smaller field of view would result in $\sim$50 visible stars per pointing. To probe our 3$\sigma$ Poisson upper limit, SKA1-LOW would need to observe $\sim$14 pointings for 8\,hours each. Observing a further 450 pointings would probe the maximum likelihood power law regime, either leading to more detections or placing stronger constraints on the Type~II burst rate. 

\section{Conclusions and future work} \label{sec:conc}
We have conducted the largest unbiased search for stellar Type~II radio bursts, targeting all stars within 100\,pc in LoTSS; this resulted in a cumulative total of 107.21\,stellar years of observation time. Our search for drifting stellar radio bursts led to the detection of two events: the previously published 2-minute burst from the M~dwarf StKM1-1262 and a new 13-minute burst from the M~dwarf LP215-56. These bursts exhibit drifts of $-0.801^{+0.003}_{-0.003}$\,MHz\,s$^{-1}$ and $-0.060^{+0.002}_{-0.002}$\,MHz\,s$^{-1}$, respectively, and both are significantly circularly polarised ($>$\,50 per cent). Given their characteristics, we assumed that both bursts were produced by plasma emission that was likely tracing stellar CMEs. Additionally, we identified 19 non-drifting radio bursts, each originating from a different star, which warrant further investigation in future studies.

Using the two drifting burst detections, we calculated Poisson upper and lower limits for the burst rate of drifting stellar radio bursts, which can be interpreted as the occurrence rate of stellar CMEs. We also fitted our data to a cumulative luminosity distribution, obtaining a power law index ($\alpha$) of -$0.7^{+0.9}_{-0.6}$. This result is consistent with decametric solar Type~II observations, where $\alpha = -0.81 \pm 0.06 \pm 0.02$. A natural next step is to compare our findings with LOFAR-detected metric solar radio bursts, which would provide a more robust comparison between stellar and solar Type~II bursts.

We find that the rarity of drifting stellar radio bursts is comparable to that of high-energy stellar flares on M~dwarfs ($\sim$10$^{35}$\,ergs). However, this comparison is only based on event rates, and future work to establish a direct physical link between optical flares and Type II bursts will require substantial co-observing in the two bands. For instance, for the Alpha Centauri~A star, SKA-Low phase-1 will be sensitive to bursts with spectral energy $\geq 10^{12}$\,erg\,s$^{-1}$\,Hz$^{-1}$, which should occur once every 275\,hours. Co-observing with a large survey that simultaneously observes multiple stars would also be fruitful. For example, an SKA survey similar to LoTSS, consisting of\,500 pointings each lasting 8\,hours, could effectively probe the maximum likelihood regime for energies $\geq10^{15}$\,erg\,s$^{-1}$\,Hz$^{-1}$ and explore the burst rate regime derived in this study.

\begin{acknowledgements}
The LOFAR data in this manuscript were (partly) processed by the LOFAR Two-Metre Sky Survey (LoTSS) team. This team made use of the LOFAR direction independent calibration pipeline (\url{https://github.com/lofar-astron/prefactor}), which was deployed by the LOFAR e-infragroup on the Dutch National Grid infrastructure with support of the SURF Co-operative through grants e-infra 170194 e-infra 180169 \citep{2017isgc.confE...2M}. The LoTSS direction dependent calibration and imaging pipeline (\url{http://github.com/mhardcastle/ddf-pipeline/}) was run on compute clusters at Leiden Observatory and the University of Hertfordshire, which are supported by a European Research Council Advanced Grant [NEWCLUSTERS-321271] and the UK Science and Technology Funding Council [ST/P000096/1]. The J\"ulich LOFAR Long Term Archive and the German LOFAR network are both coordinated and operated by the J\"ulich Supercomputing Centre (JSC), and computing resources on the supercomputer JUWELS at JSC were provided by the Gauss Centre for Supercomputing e.V. (grant CHTB00) through the John von Neumann Institute for Computing (NIC).

DCK, HKV and EI acknowledge funding from the ERC starting grant `Stormchaser' (grant number 101042416). HKV and SB acknowledge funding from the Dutch research council (NWO) under the talent programme (Vidi grant VI.Vidi.203.093). 

JRC acknowledges funding from the European Union via the European Research Council (ERC) grant Epaphus (project number: 101166008).

MJH thanks STFC for support [ST/Y001249/1].

AD acknowledges support by the BMBF Verbundforschung under the grant
05A23STA.

PZ acknowledges funding from the European Research Council (ERC) under the European Union’s Horizon 2020 research and innovation programme (grant agreement No 101020459 - Exoradio).
\end{acknowledgements}

\section*{Data Availability}

The relevant code and data products for this work will be uploaded on Zenodo at the time of publication.

\bibliographystyle{aa} 
\bibliography{references.bib}

\clearpage

\appendix
\section{Faraday rotation of the burst} \label{appendix:faraday}
We calculated the linear polarisation flux density across various rotation measure (RM) trials to determine whether burst B21 contains a linearly polarised emission component. The Faraday effect causes the plane of linear polarisation to rotate, with the degree of rotation proportional to the line-of-sight component of the magnetic field and the square of the wavelength. As a result, the linearly polarised signal oscillates sinusoidally as a function of frequency. 

\begin{figure}
    \centering
    \includegraphics[width=1\linewidth]{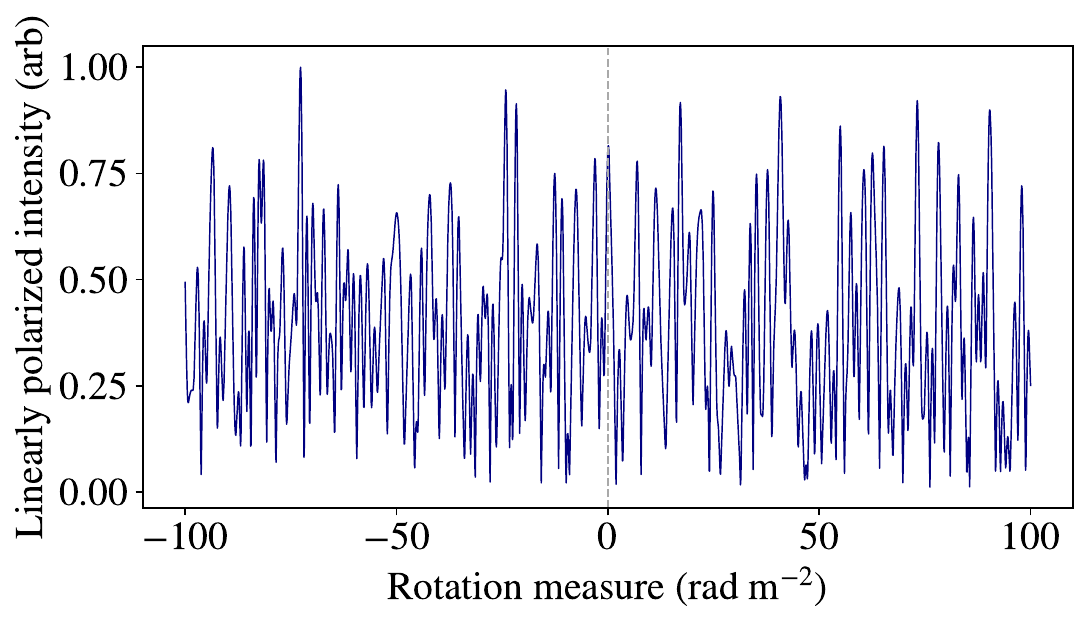}
    \caption{Linearly polarised emission of burst B21 as a function of RM trials. The absence of a significant peak indicates little to no linear polarisation in the emission.}
    \label{fig:RM}
\end{figure}

By combining the Stokes~U and Stokes~Q emission, we could compute the total linear polarisation flux density for each RM trial. By testing different RM values, we could find the polarised emission. As shown in Fig.~\ref{fig:RM} for burst B21, no significant peak of excess emission is observed at any specific RM, indicating that the burst exhibits little to no linearly polarised emission.

\section{Parker wind model} \label{appendix:parker}
We modelled the stellar wind as an isothermal Parker wind model under the assumption of a steady, spherically symmetric outflow. We considered a star of mass $M$ and radius $R$, and defined the wind properties using a mean molecular weight $\mu$, an initial magnetic field strength $B$, a temperature $T$, and an initial number density $n_0$.

The wind velocity $v$ is derived from the Parker wind solution, which satisfies the equation\begin{equation}
\left( \frac{v}{a} \right)^2 - \ln \left( \frac{v}{a} \right)^2 = 4 \ln \left( \frac{r}{r_c} \right) + C,
\end{equation}
where $a$ is the isothermal sound speed and $r_c$ is the critical point given by\begin{equation}
r_c = \frac{G M}{2 a^2}.
\end{equation}
We obtained the velocity profile, $v (r)$, numerically and extract the wind speed at the stellar surface, $v_0$. The number density, $n(r),$ follows from mass conservation:
\begin{equation}
n(r) = n_0 \frac{R^2 v_0}{r^2 v(r)}.
\end{equation}
We assumed a dipolar field that decays as $r^{-3}$ until the Alfv\'{e}n radius, where the ram pressure equals the magnetic pressure:
\begin{equation}
B(r) = B_0 \left( \frac{r}{R} \right)^{-3}, \quad \text{for } r < R_A,
\end{equation}
\begin{equation}
B(r) = B(R_A) \left( \frac{r}{R_A} \right)^{-2}, \quad \text{for } r \geq R_A.
\end{equation}
With this, we could calculate the Alfv\'{e}n speed:
\begin{equation}
v_A = \frac{B}{\sqrt{4\pi\,n\,m_p}}.
\end{equation}

The speed of the shock must exceed the wind speed in order to be a shock and must exceed the Alfv\'{e}n speed to produce the instabilities required for radio emission. Using the density profile, we calculated the plasma frequency and determined the shock speed at the moment it passes through our LOFAR observing band (i.e. 120\,MHz and 168\,MHz) and at what distance $r$ this happens. With this, we determined the drift delay of the burst:

\begin{equation}
\Delta t = \frac{r_2 - r_1}{(v_{\text{shock,1}} + v_{\text{shock,2}})/2}.
\end{equation}

\label{LastPage}
\end{document}